\documentclass[pra,twocolumn,floatfix,superscriptaddress,longbibliography, nofootinbib]{revtex4-1}
\usepackage{amsmath,amsfonts,amssymb,amsthm,graphics,graphicx,epsfig,bbm}
\usepackage[colorlinks=true,citecolor=blue,linkcolor=blue,urlcolor=blue]{hyperref}
\usepackage[usenames]{color}
\usepackage[dvipsnames]{xcolor}
\usepackage{graphicx}
\usepackage{tikz}
\usepackage{subfigure}
\usepackage{float}
\usepackage{amsmath}
\usepackage{epsfig}
\usepackage{dcolumn}
\usepackage{bm}
\usepackage{color}
\usepackage{times}
\usepackage{epstopdf}
\usepackage{amscd}
\usepackage{amsthm}
\usepackage{amssymb}
\usepackage{bbm}
\usepackage{amssymb}
\usepackage{amstext}
\usepackage{latexsym}
\usepackage{comment}
\usepackage{hyperref}
\usepackage{amsfonts}
\usepackage{psfrag}
\usepackage{soul,xcolor}
\usepackage[normalem]{ulem}
\usepackage{dsfont}
\usepackage{txfonts}
\usepackage{footnote}
\usepackage{multirow}
\usepackage{appendix}
\usepackage{mathtools}
\usepackage{bbold}
\usepackage{enumitem}
\usepackage{xstring}
\usepackage{cancel}
\usepackage{amscd}
\usepackage{bbm}
\usepackage{units}
\usepackage{cancel}
\usepackage{xstring}
\usepackage{import}
\usepackage{subfiles}
\usepackage{amsmath,amsfonts,amssymb,amsthm,graphics,graphicx,epsfig,bbm}
\usepackage[colorlinks=true,citecolor=blue,linkcolor=blue,urlcolor=blue]{hyperref}
\usepackage[usenames]{color}
\usepackage{graphicx}
\usepackage{subfigure}
\usepackage{amsmath}
\usepackage{epsfig}
\usepackage{dcolumn}
\usepackage{bm}
\usepackage{color}
\usepackage{times}
\usepackage{epstopdf}
\usepackage{amssymb}
\usepackage{amstext}
\usepackage{latexsym}
\usepackage{hyperref}
\usepackage{amsfonts}
\usepackage{psfrag}
\usepackage{soul,xcolor}
\usepackage[normalem]{ulem}
\usepackage{dsfont}
\usepackage{txfonts}
\usepackage{xcolor}
\usepackage{footnote}
\usepackage{multirow}
\usepackage{appendix}
\usepackage{mathtools}
\usepackage{ulem}
\usepackage{mathtools}

\definecolor{mycolor}{RGB}{106,81,162}

\usepackage{mathtools}

\newcommand{\iden}[1]{
    \ifthenelse{\equal{1}{\string #1}}
  {
   \mathbbm{1}
  }
  {
   \mathbbm{1}^{\otimes#1}}
  }
\newcommand{\ketzero}[1]{
    \ifthenelse{\equal{1}{\string #1}}
  {
   \ket{0}
  }
  {
   \ket{0}^{\otimes#1}}
  }
\newcommand{\brazero}[1]{
    \ifthenelse{\equal{1}{\string #1}}
  {
   \bra{0}
  }
  {
   \bra{0}^{\otimes#1}}
  }
\newcommand{\ketone}[1]{
      \ifthenelse{\equal{1}{\string #1}}
    {
     \ket{1}
    }
    {
     \ket{1}^{\otimes#1}}
    }
  \newcommand{\braone}[1]{
      \ifthenelse{\equal{1}{\string #1}}
    {
     \bra{1}
    }
    {
     \bra{1}^{\otimes#1}}
    }

\usepackage{todonotes}

\usepackage{physics}  
\usepackage{dsfont}

\begin{document}

\title{Neural quantum kernels: training quantum kernels with quantum neural networks}

\author{Pablo Rodriguez-Grasa}
\affiliation{Department of Physical Chemistry, University of the Basque Country UPV/EHU, Apartado 644, 48080 Bilbao, Spain}
\affiliation{EHU Quantum Center, University of the Basque Country UPV/EHU, Apartado 644, 48080 Bilbao, Spain}
\affiliation{TECNALIA, Basque Research and Technology Alliance (BRTA), 48160 Derio, Spain}
\email[Corresponding author: ]{\qquad pablojesus.rodriguez@ehu.eus}

\author{Yue Ban}
\affiliation{Instituto de Ciencia de Materiales de Madrid (CSIC), Cantoblanco, E-28049 Madrid, Spain}
\affiliation{Departamento de F\'isica, Universidad Carlos III de Madrid, Avda. de la Universidad 30, 28911 Legan\'es, Spain}
\affiliation{TECNALIA, Basque Research and Technology Alliance (BRTA), 48160 Derio, Spain}

\author{Mikel Sanz}
\affiliation{Department of Physical Chemistry, University of the Basque Country UPV/EHU, Apartado 644, 48080 Bilbao, Spain}
\affiliation{EHU Quantum Center, University of the Basque Country UPV/EHU, Apartado 644, 48080 Bilbao, Spain}
\affiliation{IKERBASQUE, Basque Foundation for Science, Plaza Euskadi 5, 48009, Bilbao, Spain}
\affiliation{Basque Center for Applied Mathematics (BCAM), Alameda de Mazarredo, 14, 48009 Bilbao, Spain}
\date{\today}

\begin{abstract}
Quantum and classical machine learning have been naturally connected through kernel methods, which have also served as proof-of-concept for quantum advantage. Quantum embeddings encode classical data into quantum feature states, enabling the construction of embedding quantum kernels (EQKs) by measuring vector similarities and projected quantum kernels (PQKs) through projections of these states. However, in both approaches, the model is influenced by the choice of the embedding. In this work, we propose using the training of a quantum neural network (QNN) to construct neural quantum kernels: both neural EQKs and neural PQKs, which are problem-inspired kernels. Unlike previous methods in the literature, our approach requires the kernel matrix to be constructed only once. We present several strategies for constructing neural quantum kernels and propose a scalable method to train an $n$-qubit data re-uploading quantum neural network (QNN). We provide numerical evidence of the performance of these models under noisy conditions. Additionally, we demonstrate how neural quantum kernels can alleviate exponential concentration and enhance generalization capabilities compared to problem-agnostic kernels, positioning them as a scalable and robust solution for quantum machine learning applications.
\end{abstract}

\maketitle

\section{Introduction}
Quantum computing is a promising computational paradigm for tackling some complex computational problems which are classically intractable. In particular, its potential in enhancing machine learning tasks has garnered significant attention \cite{Biamonte_2017,Carleo_2019,Dunjko_2018,dalzell2023quantum} with parametrized quantum circuits as the most common approach \cite{Benedetti_2019,Skolik_2022,Schuld_centric,farhi2018classification}. Although there are evidences of quantum advantage in some tailored problems \cite{Sweke_2021, Jerbi_2021, Pirnay_2023, gyurik2023establishing}, advantage over classical counterparts for practical applications remains as an area of active research.

Amidst the exploration of quantum machine learning models, previous studies, particularly in Refs.~\cite{enhanced_feature, schuld2019quantum}, have delineated a categorization into explicit and implicit models. In explicit models, data undergoes encoding into a quantum state, and then a parametrized measurement is performed. In contrast, implicit or kernel models are based on a weighted summation of inner products between encoded data points. A specialized category within parametrized quantum circuits, which can be considered separately, consists of data re-uploading models \cite{perez2020data}. This architecture features an alternation between encoding and processing unitaries, yielding expressive models that have found extensive use \cite{Schuld_Johannes, Caro_2021, Ono_2023}. However, Ref.~\cite{jerbi2023quantum} shows that these models can be mapped onto an explicit model, thereby unifying these three approaches within the framework of linear models within Hilbert space.

Quantum kernel methods have garnered significant attention, both for evaluating their performance \cite{Huang_2021, peters2021machine, Bartkiewicz_2020, Kusumoto_2021, Wu2023quantumphase, kyriienko2022unsupervisedquantummachinelearning} and for theorizing their role in explaining quantum machine learning models \cite{schuld2019quantum,schuld2021supervised,schuldbook}. This interest stems from several key factors. First, embedding data into quantum states provides access to the exponentially large Hilbert space, enabling efficient computation of inner products. Second, quantum feature maps facilitate the construction of quantum kernels, which may exhibit classical intractability and hold promise for quantum advantage \cite{rigorous}. Third, kernel methods, unlike neural networks, solve learning tasks through convex optimization in high-dimensional feature spaces, ensuring optimal solutions. This aligns with the representer theorem \cite{book}, which guarantees that kernel methods achieve a training loss lower than or equal to explicit models, given the same encoding and dataset. However, this enhanced expressivity can sometimes compromise generalization, as discussed in Ref.~\cite{jerbi2023quantum}.

However, quantum kernel methods present several challenges. Firstly, the computational complexity of constructing the kernel matrix scales quadratically with the number of training samples. Additionally, selecting the appropriate embedding that defines the kernel function is problem-dependent and requires careful consideration \cite{Blank2020, salmenperä2024impactfeatureembeddingplacement, quantum_tangent_kernel}. As highlighted in Ref.~\cite{kübler2021inductive}, building meaningful quantum machine learning models necessitates incorporating problem-specific knowledge. This can include, for instance, exploiting information about symmetries \cite{Wang_2025, Belis_2024, Glick_2024, exploiting_sym, group_invariant}, among other data properties. This approach helps mitigate a key issue inherent in problem-agnostic methods: the exponential concentration of kernel values \cite{Thanasilp2024}, which is particularly pronounced in embedding quantum kernels (EQKs), where kernels are constructed by calculating inner products between large quantum states. In this context, a different class of quantum kernels known as projected quantum kernels (PQKs) \cite{Huang_2021} generates kernels through projections rather than full inner product calculations, thus alleviating the exponential concentration problem. However, an effective strategy is still required to design the appropriate embedding for these kernels.

When prior information about the problem is unavailable for designing a well-tuned embedding, problem-informed embeddings can be constructed through optimization. Two main approaches have been considered for this purpose: multiple kernel learning and kernel target alignment. In the former, a combination of kernel functions is optimized to minimize an empirical risk function \cite{vedaie2020quantum, ghukasyan2023quantumclassical}. In the latter, a kernel is defined using an embedding ansatz that is trained to align with an ideal kernel \cite{Hubregtsen_2022}. However, both methods necessitate recomputing the kernel matrix at each training iteration, leading to significant computational costs.

In this article, we propose the use of a quantum neural network (QNN) to construct neural quantum kernels. This approach allows for training the embedding that generates the kernel, requiring the kernel matrix to be constructed only once. Since training a QNN is not a straightforward task, we introduce a novel approach to scale the training of a data re-uploading QNN to $n$ qubits. This construction which has been already tested in a real-world classification problem \cite{Rodriguez-Grasa_2025}, aims to circumvent trainability issues and clarifies the role of entanglement in accuracy, a factor that was previously unclear in most methods \cite{bowles2024betterclassicalsubtleart}. We demonstrate how our approach can be applied to both EQKs, identifying two specific cases: the $1$-to-$n$ approach, where a single-qubit neural network constructs an EQK for $n$ qubits, and the $n$-to-$n$ approach, where the neural network directly serves as the embedding to build the kernel, as well as to PQKs, thus providing broad applicability. Through numerical experiments, we demonstrate the effectiveness of our proposal by assessing not only model performance but also trainability and generalization ability, highlighting the importance of pre-training and the robustness of our method. Additionally, we evaluate its performance under noisy conditions.

\section{Quantum kernel methods}
Kernel methods are machine learning models defined by the linear combination
\begin{equation}\label{implicit}
    f_{\boldsymbol{\alpha},X}(\boldsymbol{x})=\sum_{i=1}^M \alpha_i\; k(\boldsymbol{x}, \boldsymbol{x}_i),
\end{equation}
where $\{\boldsymbol{x}_i\}_{i=1}^M$ represents training points from a set $X$. The kernel function $k(\boldsymbol{x}, \boldsymbol{x}_i)$ serves is a similarity measure, which must be symmetric and positive semi-definite. According to Mercer's Theorem, kernels can be estimated by evaluating the inner product of feature vectors in a Hilbert space $\mathcal{H}$, i.e., $k(\boldsymbol{x}_i, \boldsymbol{x}_j)=\langle \phi (\boldsymbol{x}_i), \phi(\boldsymbol{x}_j) \rangle_\mathcal{H}$. These feature vectors are obtained by applying a feature map $\phi$ to a data point $\boldsymbol{x}$. The parameters $\boldsymbol{\alpha}$ are determined by solving a convex optimization problem. While there are proposals for solving this type of optimization problem on a quantum device \cite{rebentrost2014quantum, Park_2023}, we assume it is performed on a classical computer. This requires constructing the kernel matrix $K$, where each entry is given by $k_{ij} = k(\boldsymbol{x}_i, \boldsymbol{x}_j)$. Constructing $K$ involves $\mathcal{O}(M^2)$ inner product evaluations between the feature vectors.

In quantum kernel methods, feature vectors are quantum states, and the feature space is the Hilbert space where these states reside. While other constructions, such as projected quantum kernels \cite{Huang_2021}, are possible, most quantum kernel constructions focus on embedding quantum kernels (EQKs), which are proven to be universal \cite{Gil-Fuster_2024}. EQKs are defined by using a quantum embedding that encodes data into quantum states $\rho(\boldsymbol{x})$, implicitly defining the EQK
\begin{equation}
    k_{ij}=\mathrm{tr}(\rho(\boldsymbol{x}_i)\;\rho(\boldsymbol{x}_j)).
\end{equation}
For pure states, the quantum embedding $S(\cdot)$ defines the quantum feature states on $n$ qubits as $\rho(\boldsymbol{x})=S(\boldsymbol{x})|\boldsymbol{0}\rangle \langle \boldsymbol{0}|S(\boldsymbol{x})^\dagger$, where $|\boldsymbol{0}\rangle \equiv |0\rangle^{\otimes n}$. Thus, the kernel simplifies to:
\begin{equation}\label{explicit_EQK}
    k_{ij}=|\langle \boldsymbol{0}|S(\boldsymbol{x}_i)^\dagger S(\boldsymbol{x}_j)|\boldsymbol{0}\rangle|^2.
\end{equation}
This quantity can be estimated on a quantum computer through various methods \cite{Buhrman_2001, Fanizza_2020, Cincio_2018}. One approach involves preparing the state $S^\dagger(\boldsymbol{x}_i) S(\boldsymbol{x}_j)|\boldsymbol{0}\rangle$ and measuring the probability that all qubits are in the $|0\rangle$ state.

An important insight from classical machine learning theory is captured by the representer theorem. The quantum version of this theorem asserts that, given a quantum embedding and a training set, models in the form of Eq.~\ref{implicit} consistently achieve training losses that are equal to or lower than those of explicit models, which take the form
\begin{equation}
    f_{\boldsymbol{\theta}}(\boldsymbol{x})= \mathrm{tr}(\rho(\boldsymbol{x})\; \mathcal{O}_{\boldsymbol{\theta}}),
\end{equation}
where $\mathcal{O}_{\boldsymbol{\theta}}$ is a parametrized observable that can be tuned.
\begin{figure}[t!]
\centering
\includegraphics[width=\columnwidth]{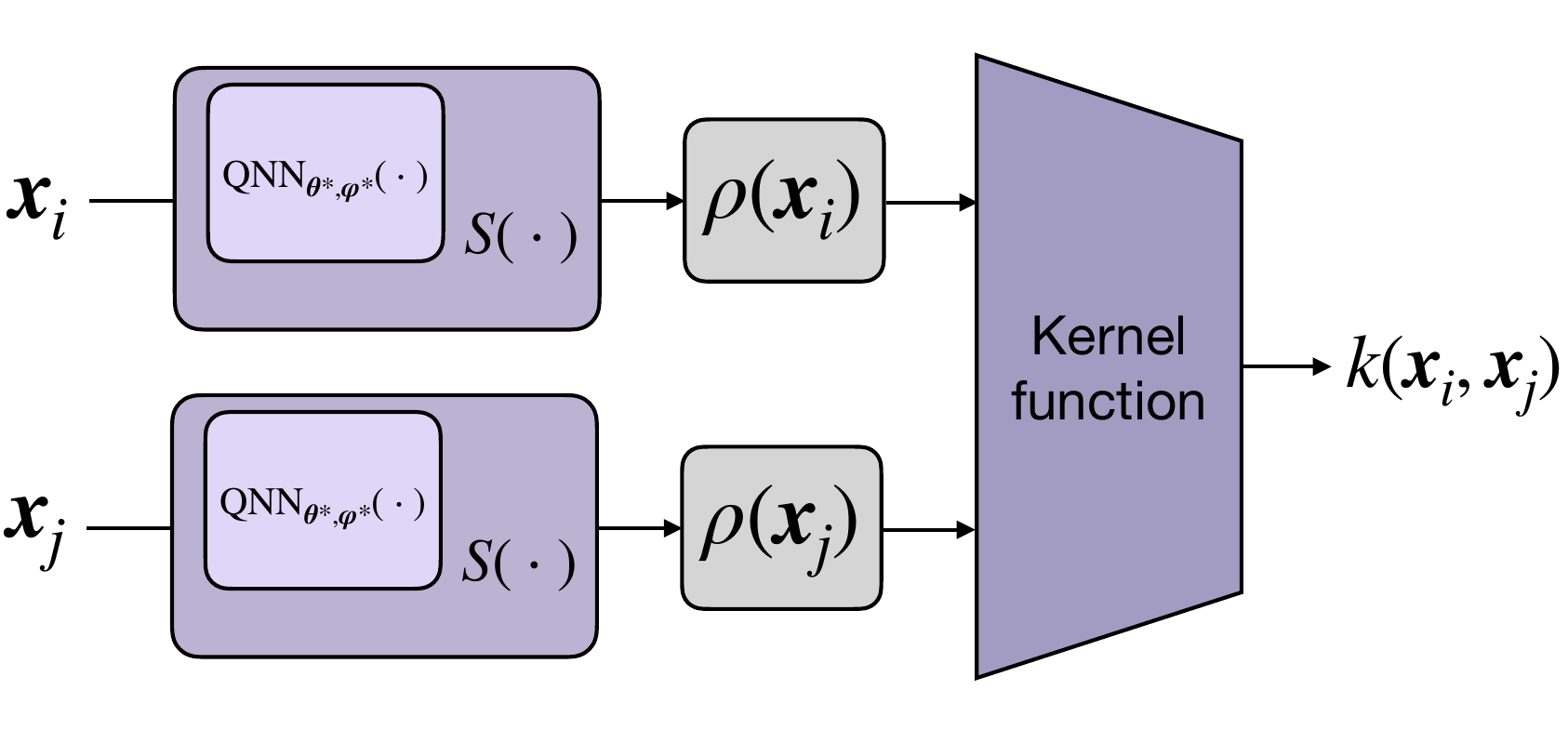}
\caption{Illustration of neural quantum kernels. Using a trained quantum neural network $\mathrm{QNN}_{\boldsymbol{\theta}^*,\boldsymbol{\varphi}^*}(\cdot)$ we construct a quantum embedding $S(\cdot)$ which generates quantum feature states $\rho(\cdot)$. By applying a kernel function to pairs of these quantum feature states, we obtain the quantum kernel $k(\boldsymbol{x}_i,\boldsymbol{x}_j)$. Depending on the nature of this kernel function, we can construct either neural EQKs or neural PQKs.}
\label{Fig:basic_protocol}
\end{figure}
Given the representer theorem, the proof of concept that quantum embeddings can lead to quantum advantages \cite{rigorous}, and the inherent capability of quantum devices to access exponentially large feature spaces, quantum kernel methods emerge as promising candidates for quantum machine learning models \cite{schuld2021supervised}. They also offer provable guarantees and flexibility, similar to training neural networks with large hidden layers, which is equivalent to using neural tangent kernels \cite{quantum_tangent_kernel}. However, accessing such large spaces is associated with the issue of exponentially vanishing kernel values \cite{kübler2021inductive,bandwidth, canatar2023bandwidth,Thanasilp2024,Suzuki_2024}. As studied in Ref.~\cite{Thanasilp2024}, problem-inspired embeddings can mitigate this issue compared to problem-agnostic embeddings. Therefore, designing suitable quantum embeddings tailored to the problem is crucial. When information about the data, such as symmetries \cite{group_invariant, exploiting_sym, Sauvage_2024, Glick_2024}, is available, it can be leveraged to construct an appropriate quantum embedding. For example, Ref.~\cite{rigorous} demonstrates a classification task based on the discrete logarithm problem, using a quantum embedding inspired by Shor’s factorization algorithm. However, in the absence of such information, quantum kernel training becomes particularly useful.
\begin{figure}[t!]
\centering
\includegraphics[width=\columnwidth]{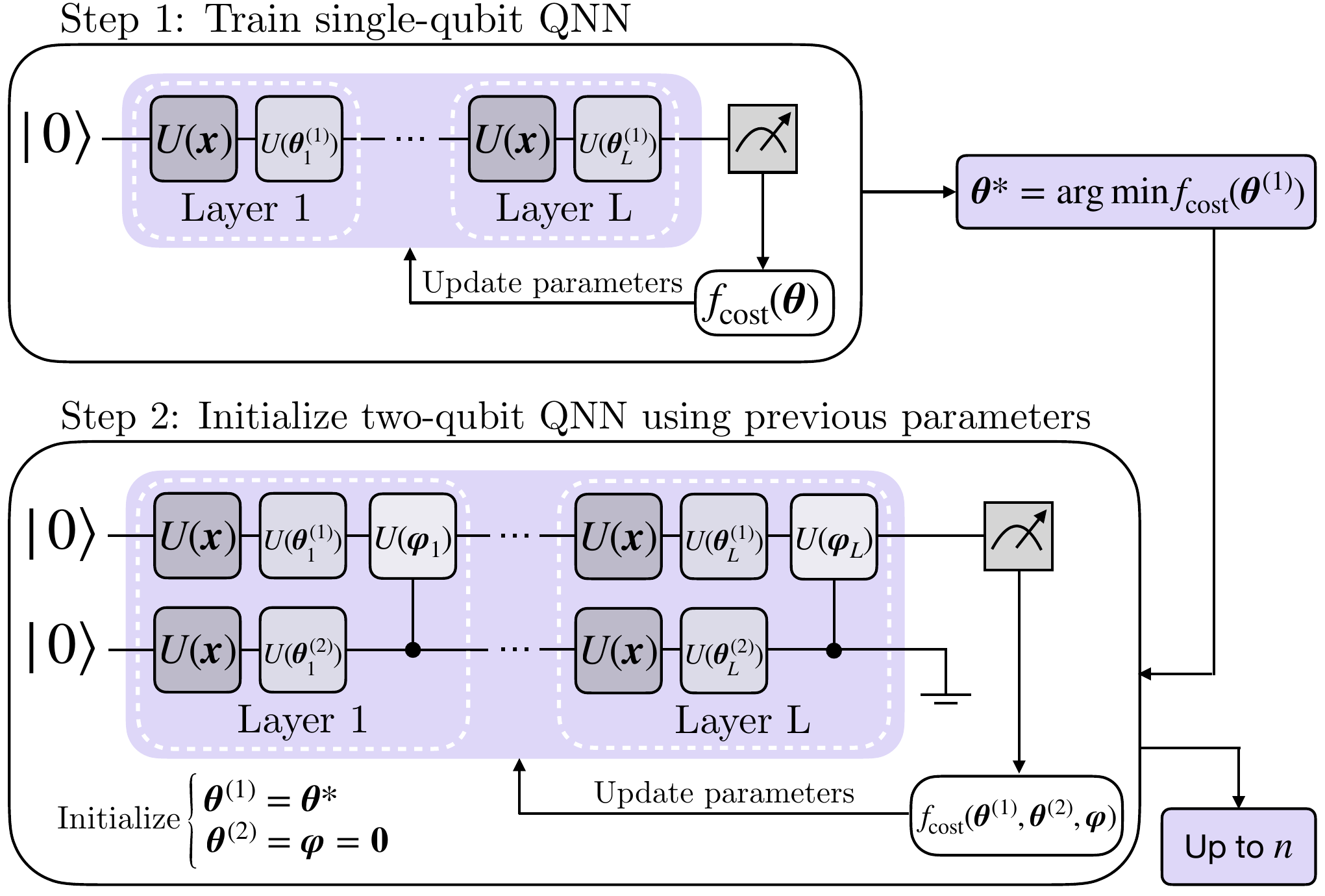}
\caption{Iterative training of a two-qubit data re-uploading QNN. In Step 1, a single-qubit QNN is trained to obtain optimal model parameters denoted as $\boldsymbol{\theta}^*$. Moving to Step 2, training for the two-qubit QNN is initiated, initializing new extra parameters to 0, while the parameters of the first qubit are set to $\boldsymbol{\theta}^*$. This iterative approach can scale the QNN up to $n$ qubits, ensuring that the $n$-qubit QNN performs at least as effectively as the $n-1$ version.}
\label{Fig:multi_qubit_QNN}
\end{figure}
The common strategy involves employing a parameterized quantum embedding $S_{\boldsymbol{\gamma}}(\cdot)$ \cite{lloyd2020quantum}, where $\boldsymbol{\gamma}$ represents the trainable parameters. This embedding defines a trainable quantum kernel $k_{\boldsymbol{\gamma}}(\boldsymbol{x}_i, \boldsymbol{x}_j)$, which is then optimized based on a specific figure of merit. For EQKs, these kernel training methods are based on multiple kernel learning, aiming to determine the optimal combination of different kernels \cite{vedaie2020quantum, ghukasyan2023quantumclassical}, and kernel target alignment \cite{Hubregtsen_2022}, where the embedding is trained to resemble an ideal kernel matrix with maximum overlap between quantum states representing the same class and minimum overlap for states of different classes. However, these approaches require constructing the kernel matrix at each training step, which implies a high computational cost. Additionally, the kernel target alignment strategy suffers from barren plateaus in the optimization \cite{Thanasilp2024}.

In our work, we propose a novel method for training embedding quantum kernels that requires constructing the kernel matrix only once. This method involves initially training a QNN and then leveraging the parameters from this training to construct the kernel. However, training a QNN is known to be challenging due to the well-documented barren plateau phenomenon \cite{McClean_2018, ragone2023unified, Holmes_2022, Cerezo_2021, Wang2021, Thanasilp2023, li2022concentrationdataencodingparameterized, larocca2024reviewbarrenplateausvariational, anschuetz2024unifiedtheoryquantumneural}. To address this, we consider a data re-uploading QNN and propose a scalable approach to use this architecture for constructing different types of quantum kernels. This strategy effectively addresses the two key issues faced by the quantum kernel alignment strategy.

\section{Scaling data re-uploading for $n-$qubit QNN}\label{sec:scaling}
As reported by Pérez-Salinas et al. in Ref.~\cite{perez2020data}, the data re-uploading model incorporates layers composed of data-encoding and training unitaries. This approach effectively introduces non-linearities to the model allowing to capture complex patterns on data \cite{Schuld_Johannes, Casas_2023, Caro_2021,barthe2023gradients}. In fact, it has been demonstrated that a single qubit quantum classifier possesses universal capabilities \cite{P_rez_Salinas_2021}.

While various encoding strategies could be considered for this architecture, we specifically adopt the easiest one defining
\begin{equation}\label{reuploading_QNN}
\begin{split}
\mathrm{QNN}_{\boldsymbol{\theta}}(\boldsymbol{x})&\equiv \prod_{l=1}^L U(\boldsymbol{\theta}_l)\;U(\boldsymbol{x})\\
&=U(\boldsymbol{\theta}_L )\;U(\boldsymbol{x}) \dots U(\boldsymbol{\theta}_1)\;U(\boldsymbol{x}).
\end{split}
\end{equation}
Here, $L$ denotes the number of layers, $U$ represents a generic $\mathrm{SU}(2)$ unitary, and the vector $\boldsymbol{\theta}=\{\boldsymbol{\theta}_1,...,\boldsymbol{\theta}_L\}$ encompasses the trainable parameters. To leverage this model to construct a binary classifier, one must select two label states that are maximally separated in Hilbert space. The training objective involves instructing the model to collectively rotate points belonging to the same class, bringing them closer to their corresponding label state.

Starting with the data re-uploading single-qubit QNN architecture, we can naturally extend it to create multi-qubit QNNs. The introduction of more qubits enhances the model's expressivity by increasing the number of trainable parameters per layer and offering the potential for entanglement \cite{panadero2023regressions}.

\begin{figure}[b!]
\centering
\includegraphics[width=\columnwidth]{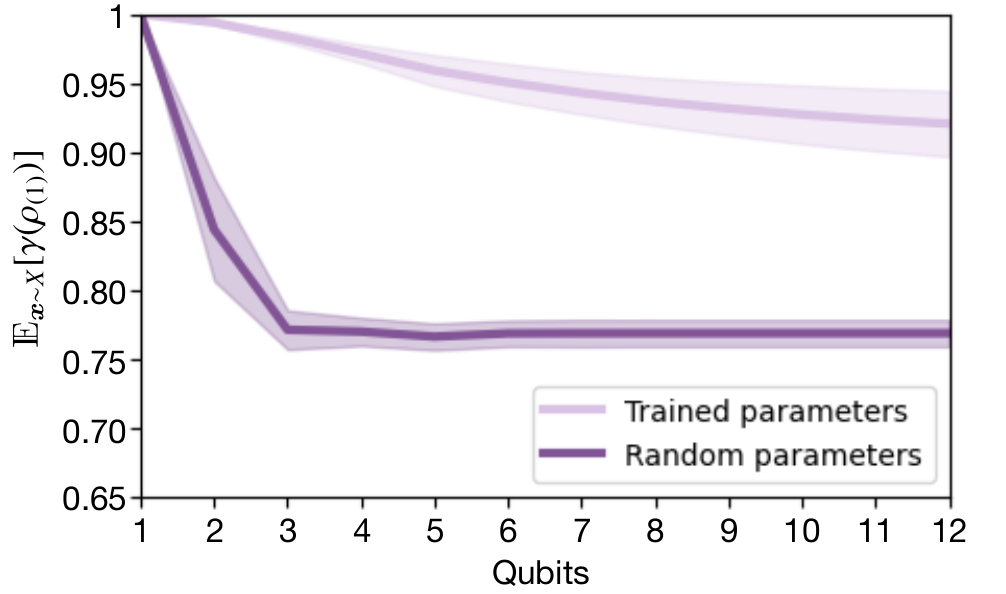}
\caption{Mean and standard deviation of the mean purity of the reduced density matrix for the first qubit, calculated over five distinct random datasets, each containing 50 data points from the Fashion MNIST dataset. The mean purity is plotted for both the proposed architecture trained iteratively and the same architecture with randomly initialized parameters.}
\label{Fig:purity}
\end{figure}

\begin{figure*}[t!]
\centering
\includegraphics[width=\textwidth]{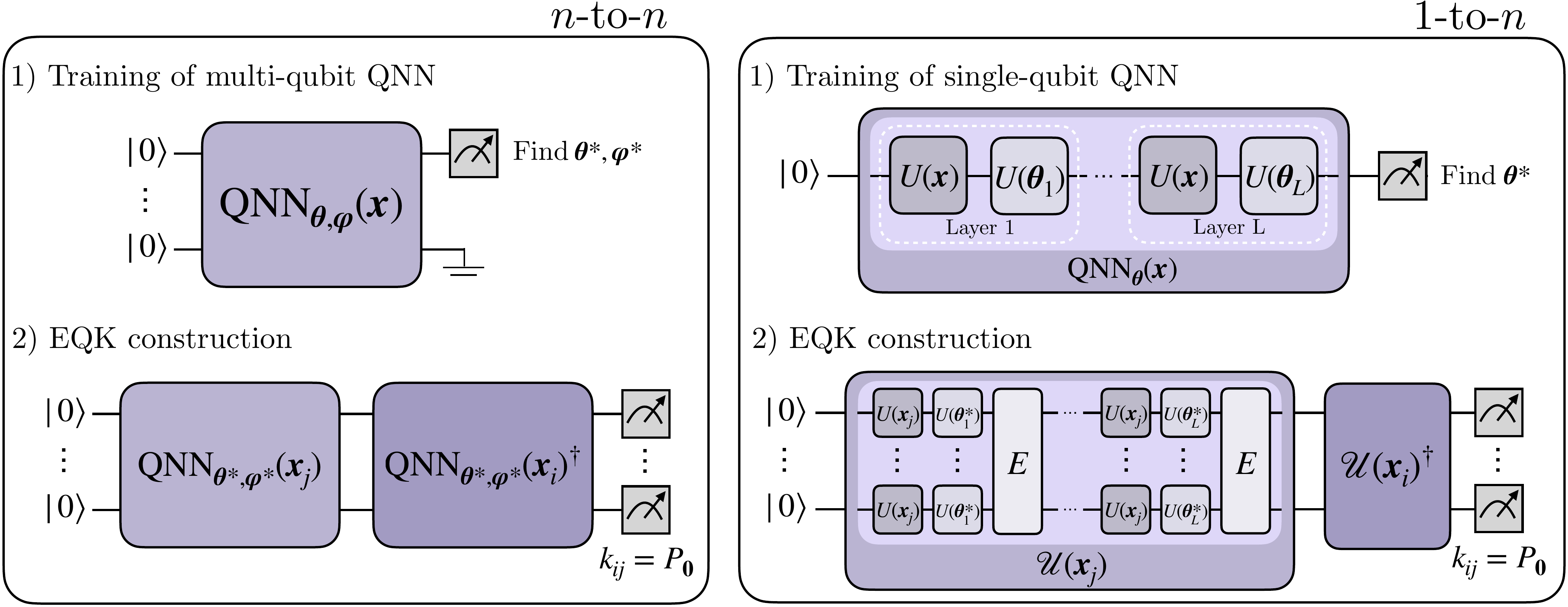}
\caption{Neural embedding quantum kernels. On the left, we have the $n$-to-$n$ proposal, constructed by directly utilizing the trained data re-uploading $n$-qubit QNN as the quantum embedding. On the right, we show the construction of an embedding quantum kernel from the training of a single-qubit QNN, named as $1$-to-$n$. The kernel matrix element $k_{ij}$ is defined as the probability of measuring all qubits in the state $|0\rangle$, denoted as $P_{\boldsymbol{0}}$.}
\label{Fig:protocol}
\end{figure*}

In this work, we propose an iterative training approach for multi-qubit QNNs. In our construction, the $n$-qubit QNN is defined as
\begin{equation}\label{multi_QNN}
\begin{split}
    \mathrm{QNN}_{\boldsymbol{\theta},\boldsymbol{\varphi}}(\boldsymbol{x})=\prod_{l=1}^L \Bigg(&\prod_{s=1}^{n-1}\mathrm{CU}^s_{s+1}(\boldsymbol{\varphi}_l^{(s)})\Big(\bigotimes_{r=1}^n U(\boldsymbol{\theta}^{(r)}_l)\Big)\;U(\boldsymbol{x})^{\otimes n}\Bigg),
\end{split}
\end{equation}
where $\prod_{i=1}^nA_i=A_n \dots  A_2 A_1$ and $\mathrm{CU}^s_{s+1}$ denotes the controlled version of the general $\mathrm{SU}(2)$ unitary with control in the $(s+1)$-th qubit and target in the $s$-th qubit, and $\boldsymbol{\theta}$ and $\boldsymbol{\varphi}$ refer to the trainable parameters of single-qubit and two-qubit gates, respectively. The total number of trainable parameters in this architecture is $3(2n-1)L$. 

To train the $n$-qubit QNN, we propose an iterative construction starting from a single-qubit QNN, as shown in Figure \ref{Fig:multi_qubit_QNN}. Initially, we train a single-qubit QNN and use its parameters to initialize the two-qubit QNN. For the two-qubit QNN initialization, we set $\boldsymbol{\varphi}^{(1)}_l=\boldsymbol{0}$ for all $l\in [1,L]$, keeping the parameters of the first qubit from the single-qubit training step. As a result, the entangling layers have no effect, so the output state of the first qubit at the beginning of the training matches that of the single-qubit QNN. This method allows us to scale the architecture, adding qubits incrementally. When adding a new qubit, we initialize the entangling gates as identities, and training starts with the optimal parameters from the previous step. This structured approach enables the systematic and scalable enhancement of the QNN's performance as more qubits are added.

Our strategy addresses key causes of barren plateaus. One major factor is the use of global cost functions; we mitigate this by using a local cost function. Another factor is entanglement: even considering a local measurement, highly entangled states can lead to barren plateaus. In our method, entanglement is introduced gradually. As shown in Figure~\ref{Fig:purity}, the resulting state remains only moderately entangled as additional qubits are added. In this figure, we plot the mean purity of the first qubit $ \gamma(\rho_{(1)}) $, defined as
\begin{equation}
    \mathds{E}_{\boldsymbol{x}\sim X}[\gamma(\rho_{(1)})]=\mathds{E}_{\boldsymbol{x}\sim X}\left[\mathrm{tr}\left(\rho^2_{(1)}(\boldsymbol{x})\right)\right]=\frac{1}{M}\sum_{i=1}^M \mathrm{tr}\left(\rho^2_{(1)}(\boldsymbol{x}_i)\right)
\end{equation}
If the state were highly entangled, the single-qubit purity would approach 1/2 , the value corresponding to a maximally mixed state. However, as shown in Figure \ref{Fig:purity}, the purity decreases only slightly with increasing system size and remains well above this value, indicating that entanglement is introduced in a structured and controlled manner. To further support this point, we compare the trained QNN to a scenario where the parameters are randomly initialized rather than optimized iteratively. In this case, the reduced state becomes more mixed, suggesting that the randomization leads to the entanglement growth and a more uniform Hilbert space exploration. The structured entanglement observed in the trained QNN helps preserve gradient magnitudes, mitigating the barren plateau problem. To generate the figure, we considered five randomly sampled datasets $X$, each containing 50 samples from the Fashion MNIST dataset.

\section{Neural embedding quantum kernels}
Considering a parameterized quantum embedding $ S_{\boldsymbol{\gamma}}(\cdot) $, in neural EQKs $ \boldsymbol{\gamma}$ are the parameters obtained from training a QNN. In the following subsections, we present two specific cases of neural EQKs: the $n$-to-$n$ and the $1$-to-$n$ configurations.

\subsection{$n$-to-$n$ approach}
In this construction, an $n$-qubit QNN is trained using the iterative method proposed in Section \ref{sec:scaling} to directly construct the corresponding EQK of $n$ qubits. As depicted in Figure \ref{Fig:protocol}, a multi-qubit QNN of the form given by Eq.~\ref{multi_QNN} is trained while fixing the trainable parameters of the embedding $\boldsymbol{\theta}^*$ and $\boldsymbol{\varphi}^*$. These parameters are then used to construct a quantum embedding
\begin{equation}
    S(\cdot)=\mathrm{QNN}_{\boldsymbol{\theta}^*,\boldsymbol{\varphi}^*}(\cdot),
\end{equation}
which defines the corresponding EQK as given by Eq.~\ref{explicit_EQK}.

This method allows us to scale the QNN as much as possible during training. Once it reaches a performance plateau, we can utilize the trained feature map to construct an EQK.

Using the representer theorem it can be shown that the kernels derived from the QNN will be at least as effective as the corresponding QNN. In our construction, the $n$-qubit QNN is defined in Eq.~\ref{multi_QNN}. There, we can extract the last layer to separate it into a variational part, which will be absorbed into the measurement when we define the corresponding quantum model, and the encoding part which will define the quantum feature map for the construction of the kernel. This corresponds to
\begin{equation}
\begin{split}
    \mathrm{QNN}_{\boldsymbol{\theta},\boldsymbol{\varphi}}(\boldsymbol{x})=\underbrace{\prod_{s=1}^{n-1}\mathrm{CU}^s_{s+1}(\boldsymbol{\lambda}^{(s)})\;\left(\bigotimes_{r=1}^n U(\boldsymbol{\omega}^{(r)})\right)}_{=V(\boldsymbol{\lambda},\boldsymbol{\omega})}\\
    \cdot\underbrace{\prod_{l=1}^{L-1} \left(\prod_{s=1}^{n-1}\mathrm{CU}^s_{s+1}(\boldsymbol{\varphi}_l^{(s)})\;\left(\bigotimes_{r=1}^n U(\boldsymbol{\theta}^{(r)}_l)\right)\;U(\boldsymbol{x})^{\otimes n}\right)}_{=S_{\boldsymbol{\theta},\boldsymbol{\varphi}}(\boldsymbol{x})},
\end{split}
\end{equation}
where we defined $\boldsymbol{\lambda}\equiv\boldsymbol{\varphi}_L$ and $\boldsymbol{\omega}\equiv \boldsymbol{\theta}_L$, which are the parameters of the last layer. This formulation aligns with the definition of a quantum model from Ref.~\cite{schuld2021supervised},
\begin{equation}
    f(\boldsymbol{x})=\mathrm{tr}(\rho(\boldsymbol{x})\mathcal{M}).
\end{equation}
In our case, 
\begin{equation}
    \rho(\boldsymbol{x}):=\rho_{\boldsymbol{\theta},\boldsymbol{\varphi}}(\boldsymbol{x})=S_{\boldsymbol{\theta},\boldsymbol{\varphi}}(\boldsymbol{x})|0\rangle\langle 0|^{\otimes n}S_{\boldsymbol{\theta},\boldsymbol{\varphi}}(\boldsymbol{x})^\dagger,
\end{equation}
and the variational measurement 
\begin{equation}
    \mathcal{M}:=\mathcal{M}_{\boldsymbol{\lambda},\boldsymbol{\omega}}=V(\boldsymbol{\lambda},\boldsymbol{\omega})\;(\hat{\sigma}_z\otimes \mathds{1}^{(n-1)})\;V(\boldsymbol{\lambda},\boldsymbol{\omega})^\dagger.
\end{equation}
\begin{figure}[t!]
\centering
\includegraphics[width=\columnwidth]{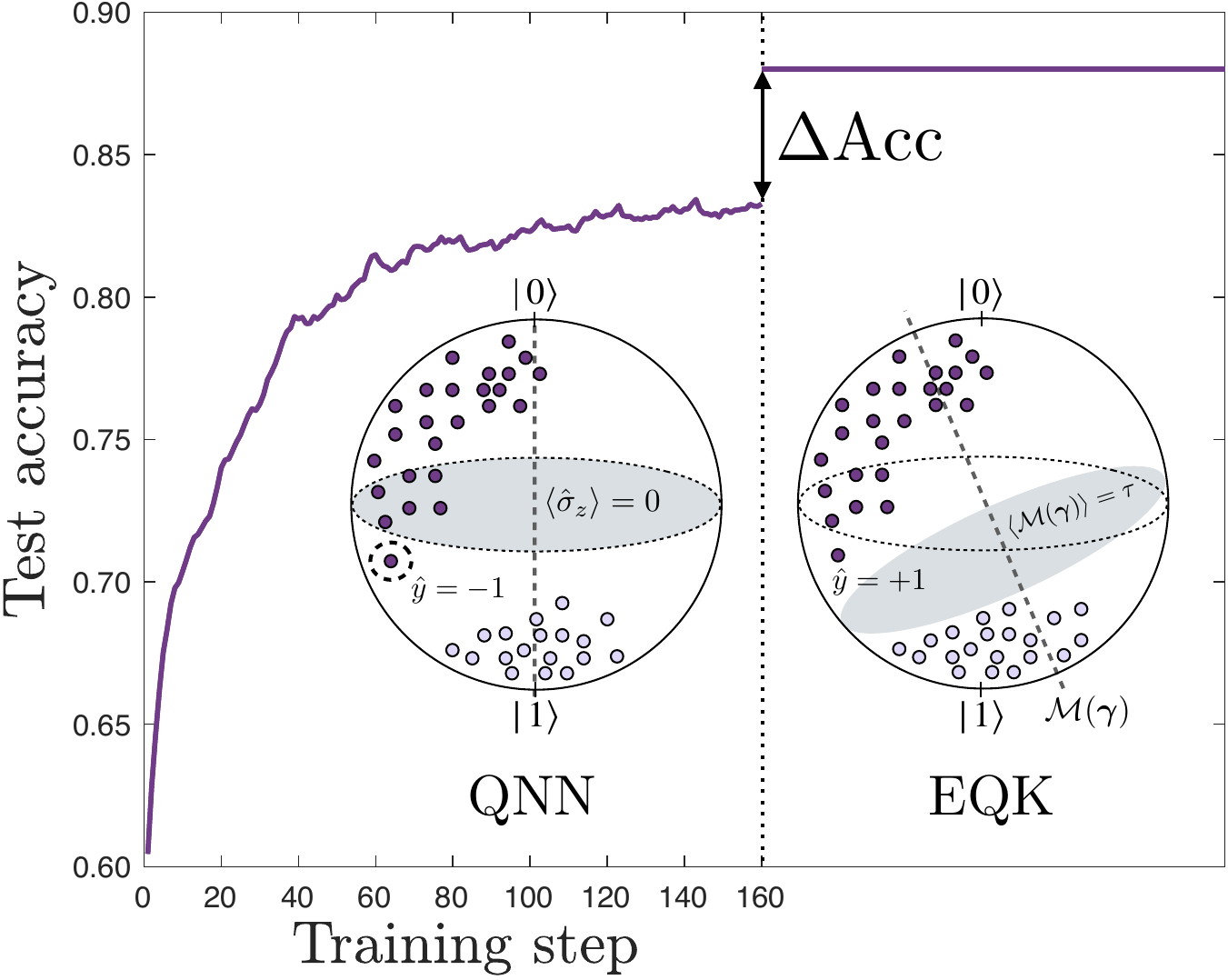}
\caption{An schematic illustration of how a single-qubit EQK, constructed from a single-qubit QNN, can enhance classification outcomes. The QNN aims to group points of the same class while fixing the decision plane (left Bloch sphere). The SVM, employing the single-qubit EQK, fine-tunes the decision boundary parameters to find the optimal hyperplane (right Bloch sphere). As a result, data points that were previously misclassified can now be correctly assigned to their respective labels. We also provide an example demonstrating a scenario in which the QNN accuracy plateaus out for a specific dataset. By using the corresponding EQK, we obtain however an increase in accuracy, denoted as $\Delta \mathrm{Acc}$.}
\label{Fig:1q1q}
\end{figure}
Once this model is trained over a dataset, we determine the parameters that minimize the $f_\mathrm{cost}$ defined in the previous section, fixing $\boldsymbol{\beta}^*,\boldsymbol{\theta}^*,\boldsymbol{\lambda}^*$, and $\boldsymbol{\omega}^*$. If we now use the corresponding feature map $S_{\boldsymbol{\theta}^*,\boldsymbol{\varphi}^*}$ to construct an EQK, we are effectively replacing the measurement from the optimization $\mathcal{M}_{\boldsymbol{\lambda}^*,\boldsymbol{\omega}^*}$ by the optimal measurement
\begin{equation}
    \mathcal{M}_{\mathrm{opt}}=\sum_{m=1}^M \alpha_m \; \rho_{\boldsymbol{\theta}^*,\boldsymbol{\varphi}^*}(\boldsymbol{x}_m)
\end{equation}
which, by the representer theorem, defines the optimal quantum model 
\begin{equation}
    f_{\mathrm{opt}}(\boldsymbol{x})=\sum_{m=1}^M\alpha_m \mathrm{tr}\left(\rho_{\boldsymbol{\theta}^*,\boldsymbol{\varphi}^*}(\boldsymbol{x})\;\rho_{\boldsymbol{\theta}^*,\boldsymbol{\varphi}^*}(\boldsymbol{x}_m)\right)
\end{equation}
that minimizes the regularized empirical risk function. Thus, we proved that constructing the EQK from QNN training will perform equally or better in terms of training loss than the QNN alone.

This formal relationship with the representer theorem can be intuitively seen in Figure \ref{Fig:1q1q}, specifically for a single-qubit scenario. Initially, the QNN rotates data points of the same class near their label state while maintaining the decision hyperplane fixed. This hyperplane is taken as the equator of the Bloch sphere, i.e., $\langle \hat{\sigma}_z\rangle=0$. After training the QNN, the resulting feature map is obtained by preserving the parameters acquired during training. This feature map is then utilized to construct an EQK. The subsequent application of the SVM algorithm, using this kernel, aims to identify the optimal separation hyperplane in the feature space. In this context, the optimization is equivalent to adjusting the optimal measurement $\mathcal{M}(\boldsymbol{\gamma})$ while keeping the data points fixed. 

Therefore, the QNN does not need to be perfectly trained;  even if the points of the same class do not align exactly with their corresponding label states,  the optimal measurement derived from the kernel construction can effectively separate the two classes.

\subsection{$1$-to-$n$ approach}
A more non-trivial approach is the $1$-to-$n$ construction, where we train a single-qubit $\mathrm{QNN}_{\boldsymbol{\theta}}$ from Eq.~\ref{reuploading_QNN}, fixing the $\boldsymbol{\theta}^*$ and we leverage this training to construct an EQK given by the embedding 
\begin{equation}\label{1-to-n}
\begin{split}
    S(\cdot)=\prod_{l=1}^L E\;U(\boldsymbol{\theta}^*_l)^{\otimes n}\;U(\cdot)^{\otimes n},
\end{split}
\end{equation}
where $E$ denotes an entangling operation, such as a cascade of $\mathrm{CNOT}$ or $\mathrm{CZ}$ gates, among other possibilities. It is worth noting that the training is conducted for a single-qubit QNN and does not explicitly consider entanglement. Nevertheless, we will present numerical results demonstrating that this training alone is sufficient to select parameters for constructing a customized multi-qubit EQK tailored to a specific task. Remarkably, the training of the single-qubit QNN could be executed on a classical computer, and subsequently, the derived parameters could be utilized to construct a potent kernel on a quantum computer.

Certainly, one can combine both the $n$-to-$n$ and the $1$-to-$n$ architectures and generalize it to $n$-to-$m \cdot n$, where $m$ represents some integer. In this scenario, an $n$-qubit QNN is trained and utilized to implement the same design as in the $1$-to-$n$ construction. However, in this case, each qubit of the QNN is embedded into $m$ qubits, introducing entanglement between layers.


\section{Neural projected quantum kernels}
Given a quantum embedding, instead of constructing a quantum kernel directly from the inner product of the full quantum feature states, Ref.~\cite{Huang_2021} introduced projected quantum kernels (PQKs). This approach builds the kernel using projections of these quantum feature states.

An example of such kernels is
\begin{equation}
    k_{ij}=\exp\left(-\gamma\sum_k \norm{\rho^{(k)}(\boldsymbol{x}_i)-\rho^{(k)}(\boldsymbol{x}_j)}_F^2\right),
\end{equation}
where $\rho_{(k)}(\boldsymbol{x}_i)=\mathrm{tr}_{i\neq k}\rho(\boldsymbol{x})$ represents the one-particle reduced density matrix on qubit $k$, $\gamma$ is a tunable hyperparameter, and $\norm{\cdot}_F$ denotes the Frobenius norm. However, these types of kernels still face the challenge of their performance being dependent on the choice of the quantum embedding that defines the quantum feature states $\rho(\boldsymbol{x})$.

\begin{figure}[t!]
\centering
\includegraphics[width=\columnwidth]{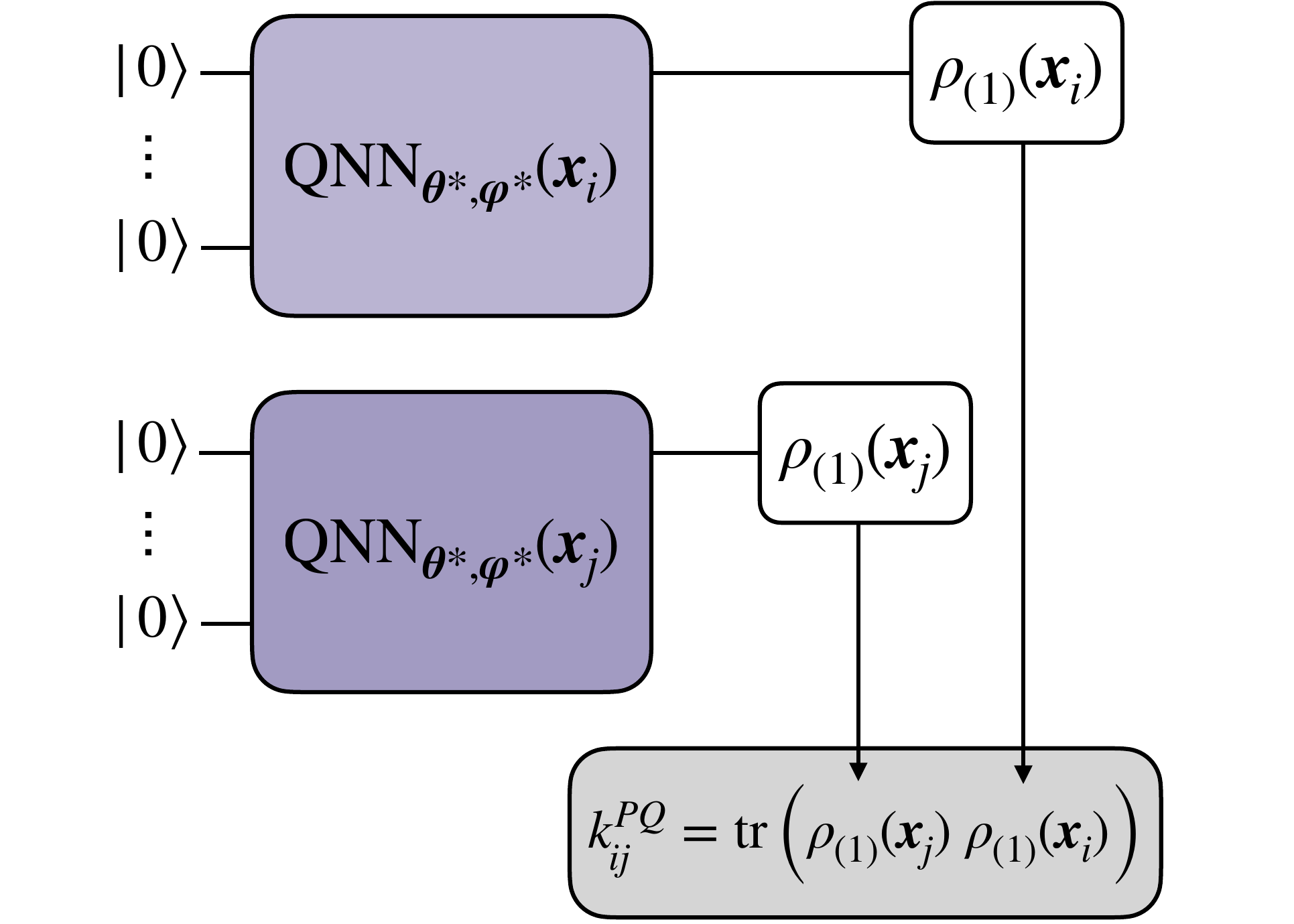}
\caption{Neural projected quantum kernels. Utilizing the trained quantum neural network $\mathrm{QNN}_{\boldsymbol{\theta}^*,\boldsymbol{\varphi}^*}(\cdot)$, we construct a projected quantum kernel by computing the trace between  reduced quantum feature states, specifically using the reduced density matrix on the first qubit $\rho_{(1)}(\cdot)$.}
\label{Fig:PQK_protocol}
\end{figure}

We can extend our proposed idea by pre-training the quantum embedding using a QNN. Specifically, we consider a simpler PQK construction given by
\begin{equation}
    k_{ij}^{PQ}=\mathrm{tr}\left(\rho_{(1)}(\boldsymbol{x}_i)\;\rho_{(1)}(\boldsymbol{x}_j)\right),
\end{equation}
where the quantum embedding is defined by a trained $n$-qubit QNN. This approach allows us to define a neural PQK by setting
\begin{equation}
    \rho_{(1)}(\boldsymbol{x})=\mathrm{tr}_{j\neq 1} \left( \mathrm{QNN}_{\boldsymbol{\theta}^*,\boldsymbol{\varphi}^*}(\boldsymbol{x})|\boldsymbol{0}\rangle\langle \boldsymbol{0}| \mathrm{QNN}_{\boldsymbol{\theta}^*,\boldsymbol{\varphi}^*}(\boldsymbol{x})^\dagger\right),
\end{equation}
which is depicted in Figure \ref{Fig:PQK_protocol}.
In Section \ref{section_numerical}, we demonstrate the importance of pre-training the parameters $\boldsymbol{\theta}$ and $\boldsymbol{\varphi}$. We compare the performance of the neural PQK with that of a PQK constructed from a randomly fixed quantum embedding, highlighting the benefits of our proposed approach.

As discussed in Ref.~\cite{Thanasilp2024}, PQKs benefit from reduced exponential concentration behaviors. Additionally, incorporating pre-training into the quantum embedding introduces an inductive bias towards the considered dataset. This bias can reduce the need for high expressivity, potentially requiring fewer layers to achieve the same performance.

\begin{figure*}[t!]
\centering
\includegraphics[width=\textwidth]{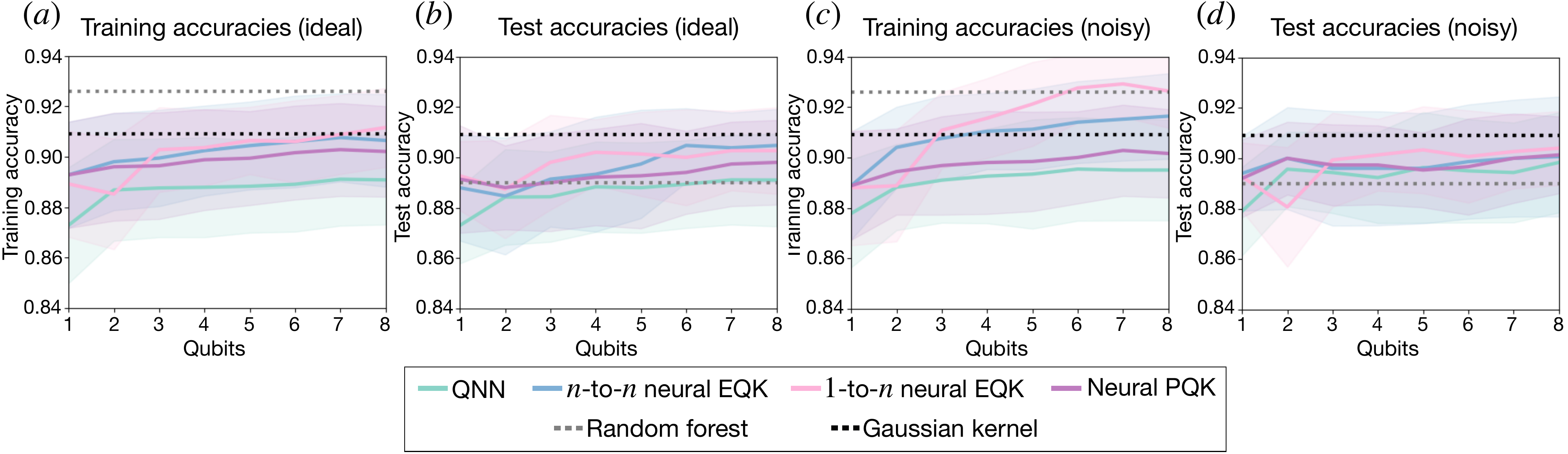}
\caption{Numerical results for the binary classification problem on the Fashion MNIST dataset. Training and test accuracies, averaged over 5 independent experiments with different random seeds, are reported for both ideal ($(a)$ and $(b)$) and noisy ($(c)$ and $(d)$) simulations. The results show the mean and standard deviation of the QNN performance using our scaling construction, as well as the three proposed neural quantum kernels.}
\label{Fig:Numerical_results_main}
\end{figure*}

\section{Numerical results}\label{section_numerical}
In this section, we present numerical results to validate the proposed neural quantum kernels. Specifically, we tackle a binary classification problem on the Fashion MNIST dataset \cite{fashion}, with the objective of distinguishing between dresses (class 3) and shirts (class 6). To reduce the dimensionality, we applied principal component analysis, compressing the feature space to 3 components—corresponding to the number of features that can be encoded using a single encoding unitary. The evaluation includes both ideal and noisy simulations, utilizing 500 training samples and 300 test samples.

The QNN architecture consists of $L=7$ layers. For training, we use the Adam optimizer with a batch size of 24. During the first training step $n=1$, the learning rate is set to 0.05, and the model is trained for 30 epochs. For subsequent steps $n>1$, the learning rate is reduced to 0.005, and training is conducted for 10 epochs.

For the noisy simulations, we applied two single-qubit quantum channels: amplitude damping and phase damping, each following the application of every quantum gate. Using the Kraus decomposition to represent the action of the noise channels
\begin{equation}
    \rho\xrightarrow{}\sum_i K_i\rho K_i^\dagger,
\end{equation}
the amplitude damping channel is described by the Kraus operators
\begin{align}
    K_0&=
    \begin{pmatrix}
        1 & 0\\
        0 & \sqrt{1-\gamma}
    \end{pmatrix},\\
    K_1&=
    \begin{pmatrix}
        0 & \sqrt{\gamma} \\
        0 & 0
    \end{pmatrix},
\end{align}
where $\gamma \in [0,1]$ represents the amplitude damping probability, which can be defined as $\gamma = 1 - e^{-\Delta t / T_1}$, with $T_1$ being the thermal relaxation time and $\Delta t$ the duration of the quantum gate application. 
For the phase damping channel, the Kraus operators are
\begin{align}
    K_0&=
    \begin{pmatrix}
        1 & 0\\
        0 & \sqrt{1-\lambda}
    \end{pmatrix},\\
    K_1&=
    \begin{pmatrix}
        0 & 0 \\
        0 & \sqrt{\lambda}
    \end{pmatrix},
\end{align}
where $\lambda \in [0,1]$ is the phase damping probability, which is given by $\lambda = 1 - e^{-\Delta t / T_2}$, and $T_2$ denotes the dephasing time. 

Considering the experimental values for the noisy parameters of a superconducting quantum processor, which are $50\; \mu s \leq T_1 \leq 150 \;\mu s$, $25\; \mu s \leq T_2 \leq 75\; \mu s$, and $10 \; ns \leq \Delta t \leq 50 \; ns$, we note that the noise level increases when $T_1$ and $T_2$ decrease and $\Delta t$ increases. For our numerical experiments, we considered the worst-case scenario for these parameters: $T_1 = 50 \;\mu s$, $T_2 = 25\; \mu s$, and $\Delta t = 90 \; ns$.

The Figure \ref{Fig:Numerical_results_main} illustrates the training and test accuracies for the different neural quantum kernel models proposed under both ideal and noisy simulation scenarios. The results represent the mean accuracy across five dataset samples. For the 1-to-n approach, the entangling operation is implemented using a cascade of CNOT gates. Taking Eq.~\ref{1-to-n}, this means considering
\begin{equation}
    E=\prod_{s=1}^{n-1}\mathrm{CNOT}^{s+1}_s.
\end{equation}
Although kernel models are prone to overfitting, we observe that in this case, the training accuracies (Fig.~\ref{Fig:Numerical_results_main} $(a)$ and $(c)$) are only slightly higher than the test accuracies (Fig.~\ref{Fig:Numerical_results_main} $(b)$ and $(d)$), highlighting the notable generalization ability of the models. This disparity is more pronounced in the noisy scenario (Fig.~\ref{Fig:Numerical_results_main} $(c)$ and $(d)$), as expected, since noise reduces the model's generalization capability. In all cases, accuracies improve as the number of qubits increases, driven by the iterative construction of the QNN. However, this improvement is less pronounced for the test accuracies in the noisy scenario.  

\begin{figure}[b!]
\centering
\includegraphics[width=\columnwidth]{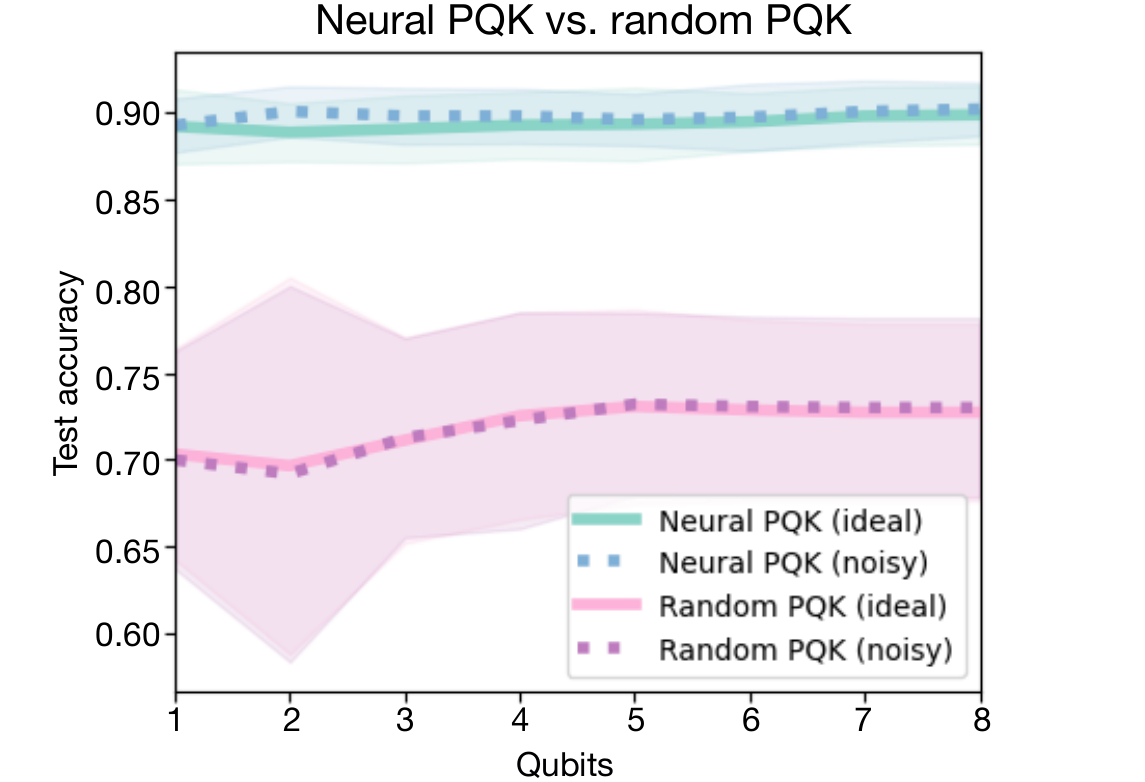}
\caption{Test accuracies for the proposed neural PQK compared to a PQK with random parameters, which is equivalent to using a problem-agnostic quantum embedding. The results are presented for both the ideal and noisy scenarios.}
\label{Fig:comparison}
\end{figure}

The QNN's performance is generally lower than that of kernel-based models. While this aligns with the representer theorem, which applies to training accuracies, we also observe that neural quantum kernels outperform the QNN in test accuracies as well. Although the QNN achieves competitive performance on its own, our proposal implies using it as a foundation for constructing quantum kernels, enabling more powerful models. However, as the kernel construction process requires additional quantum computational steps, its advantage diminishes in the presence of noise due to error propagation. Nonetheless, neural quantum kernels consistently outperform the standalone QNN, even in noisy conditions.

In the ideal scenario, especially for test accuracies, the neural EQK models tend to outperform neural PQK models. However, this performance gap narrows in the noisy scenario, as neural PQK models use circuits with half the depth of neural EQK models, making them more robust to noise.  

Furthermore, Figure~\ref{Fig:comparison} highlights the critical role of pre-training in constructing neural PQKs. It compares the test accuracies of our proposed neural PQK approach with those obtained without pre-training, where random parameters are used to create a problem-agnostic quantum embedding, referred to as random PQK. Specifically, the parameters $\boldsymbol{\theta}$ and $\boldsymbol{\varphi}$ are initialized randomly, following a uniform distribution in the range [0, 1]. The substantial difference in performance underscores the effectiveness of our neural quantum kernel strategy.

To compare the quantum models with classical methods, we employed two classical machine learning approaches as benchmarks: a support vector machine (SVM) with a Gaussian kernel and a random forest. For both methods, a grid search was conducted over the hyperparameters, and the optimal combination was selected using cross-validation.  

For the SVM model, the hyperparameters considered were the regularization parameter
\begin{equation}  
    C \in \{0.1, 1, 10, 100\},  
\end{equation}  
and the kernel hyperparameter  
\begin{equation}  
    \gamma \in \{0.01, 0.1, 1, 10\}.  
\end{equation}  

For the random forest model, the hyperparameters included the maximum depth of individual trees  
\begin{equation}  
    \text{max\_depth} \in \{2, 3, 4, 5\},  
\end{equation}  
and the number of trees
\begin{equation}  
    \text{n\_estimators} \in \{25, 50, 100, 200, 500\}.  
\end{equation}  
While the quantum models underperform these classical methods with just 1–2 qubits, increasing the number of qubits allows the quantum models to achieve competitive test accuracies.  Additional numerical results are provided in Appendix \ref{numerical_experiments}.  

These findings demonstrate two key aspects. First, the scalability of a data re-uploading QNN and the improved performance obtained by constructing the corresponding neural quantum kernel, and second the effectiveness of the $1$-to-$n$ approach in creating complex kernels by training simple QNN architectures. Although the small QNNs can be trained on classical computers, the resulting parameters can be used to construct an EQK inspired by classically hard problems, showcasing the potential for enhanced performance.

\begin{figure}[t!]
\centering
\includegraphics[width=\columnwidth]{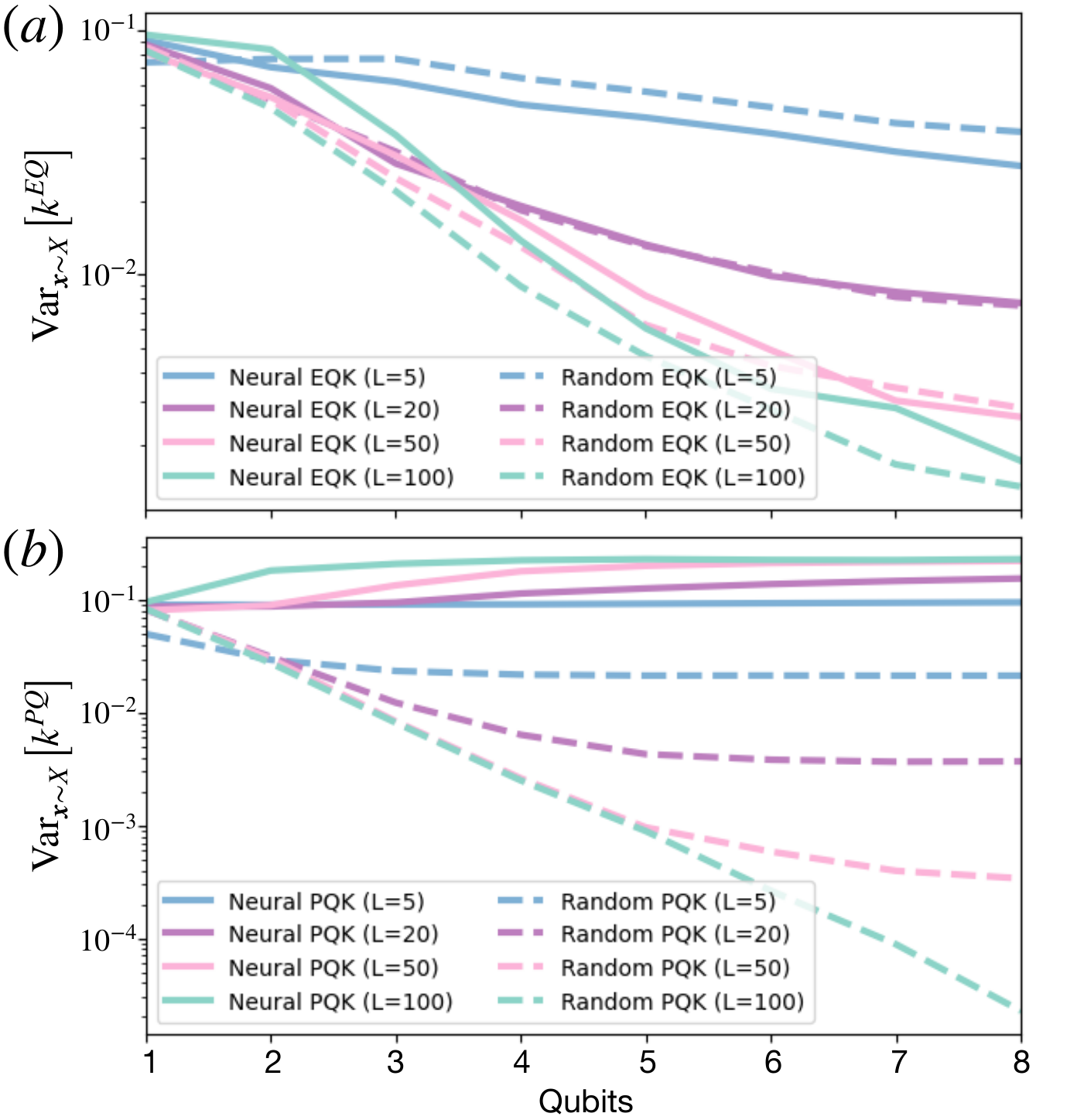}
\caption{Variances of the $(a)$ embedding and $(b)$ projected quantum kernels for both the neural construction and random parameters, plotted as a function of the number of qubits $n$ and the number of layers $L$. The data used consists of 50 samples from the Fashion MNIST dataset.}
\label{Fig:exp_concentration}
\end{figure}

\section{Trainability and generalization capacity}
To build an effective quantum machine learning model, one that can make accurate predictions on unseen data, it must be both trainable and exhibit low generalization error.

Regarding trainability, it is important to distinguish between two types: the trainability of the embedding and the trainability that follows after the embedding is selected. In our case, the first refers to the training of the QNN, which involves optimizing a non-convex cost function. The second type occurs after the embedding is chosen, where we use the kernel matrix as input and optimize the kernel model parameters by solving a convex optimization problem. In this scenario, with a fixed embedding, the optimization method is guaranteed to find the optimal solution.

However, constructing the kernel matrix $ K $ involves sampling the matrix elements 
$ k_{ij} = k(\boldsymbol{x}_i, \boldsymbol{x}_j) $ on a quantum computer. This process relies on measurements (shots), and if the differences between kernel values are very small, a large number of shots will be required to accurately estimate them. Specifically, if the kernel matrix elements exhibit exponential concentration and we only have a polynomial number of shots, the resulting optimized model may become independent of the input data, undermining its ability to generalize effectively. Using Chebyshev's inequality, the concentration of the kernel values is expressed as  
\begin{equation}  
    \mathrm{Pr}_{\boldsymbol{x}_i,\boldsymbol{x}_j} \qty[\abs{k_{ij}-\mathds{E}_{\boldsymbol{x}_i,\boldsymbol{x}_j}\qty[k_{ij}]}\geq \delta] \leq \frac{\mathrm{Var}_{\boldsymbol{x}_i,\boldsymbol{x}_j} \qty[k_{ij}]}{\delta^2}  
\end{equation}  
for any $ \delta>0 $. Here, the probability and variance are evaluated over all possible pairs of input data. This differs from the exponential concentration in QNNs, where the variance is studied in relation to the trainable parameters. Specifically, if the variance decreases exponentially with the number of qubits,  
\begin{equation}  
    \mathrm{Var}_{\boldsymbol{x}_i,\boldsymbol{x}_j} \qty[k_{ij}] \in \mathcal{O}(c^{-n}),  
\end{equation}  
for $ c > 1 $, the kernel values will concentrate exponentially around the mean. In this case, an exponential number of measurements will be required to accurately approximate the kernel matrix.

\begin{figure}[t!]
\centering
\includegraphics[width=\columnwidth]{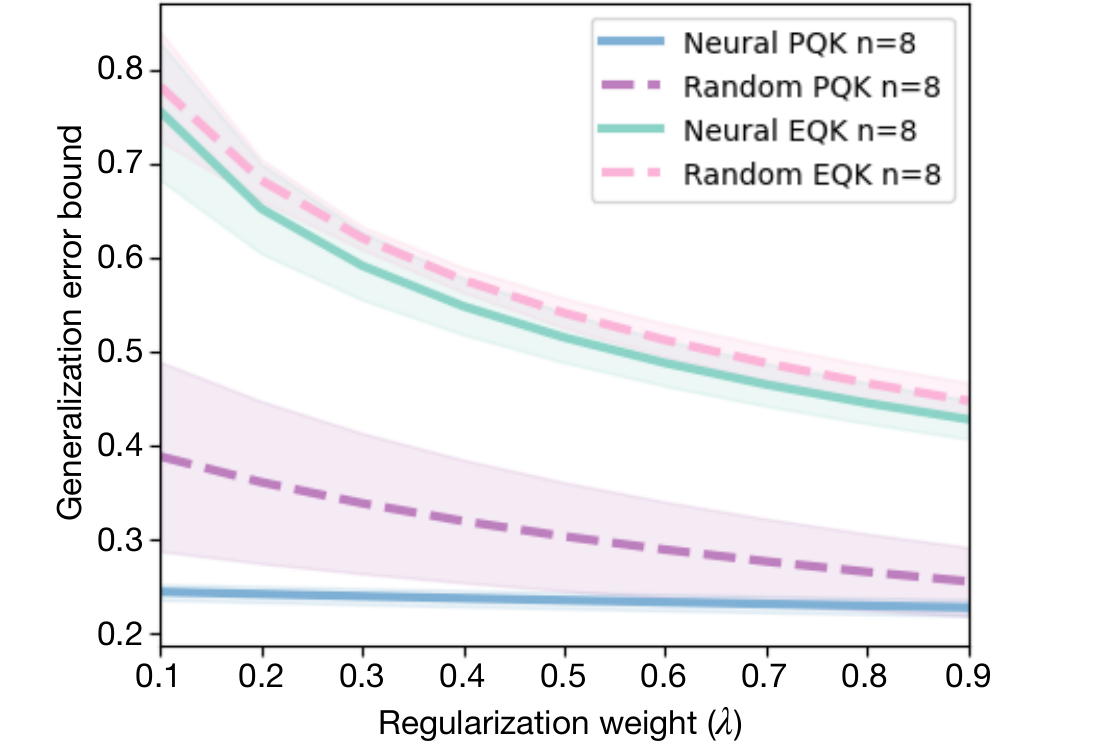}
\caption{Generalization error bound as a function of the regularization weight for both neural EQK and PQK, compared to random parameter selection for the embedding. All models use $n=8$ qubits, with the neural EQK corresponding to the $n$-to-$n$ approach. The data consists of 50 samples from the Fashion MNIST dataset.}
\label{Fig:gen_bound}
\end{figure}

Just as highly expressive quantum ansätze can lead to barren plateaus in QNNs \cite{Holmes_2022}, highly expressive quantum embeddings can result in exponential concentration in quantum kernel methods. Specifically, both embedding and projected quantum kernels exhibit this exponential decay when the ensemble of unitaries, generated by applying the embedding to the training set, is exponentially close to a 2-design (see Theorem 1 in Ref.~\cite{Thanasilp2024}). This underscores the importance of designing problem-inspired embeddings. 

In Figure~\ref{Fig:exp_concentration}, we present the scaling of the variance as a function of the number of qubits for different numbers of layers, comparing EQKs ($ k_{ij}^{EQ} $) and PQKs ($ k_{ij}^{PQ} $) in the neural construction versus problem-agnostic embeddings, where the embedding parameters are selected randomly. The variance is computed from the kernel matrix, excluding the diagonal terms for EQKs as these are equal to one. 
For EQKs, we observe that the variances for both the neural approach and random parameter selection are quite similar and do not exhibit exponential decay. In contrast, for PQKs with random parameters, the variances show an exponential decrease, particularly for $L=100$ layers. Notably, while the behavior of PQKs with random parameters aligns with previous observations in Ref.~\cite{Thanasilp2024}, the neural construction yields a strikingly different trend: the variances remain stable as the number of qubits increases. This highlights the importance of pre-training the quantum embedding to mitigate the exponential concentration of kernel values. 

On the other hand, the fact that a model is trainable does not guarantee its ability to make accurate predictions on new data, as quantum models can memorize random patterns \cite{rethinking_generalization}. Therefore, it is crucial to ensure that the model has a low generalization error, $ \epsilon_{\mathrm{gen}} $, which is defined as the difference between the true error and the training error. For a training set of $M$ data points ${(x_i,y_i)}_{i=1}^M$, this error can be upper-bounded with probability at least $ 1 - \delta $ as follows 
\begin{equation}
    \epsilon_{\mathrm{gen}} \leq \mathcal{O} \left(\sqrt{ \frac{\sum_{i=1}^M \sum_{j=1}^M A_{i,j}\;y_i y_j}{M}} + \sqrt{\frac{\log(1/\delta)}{M}} \right),
\end{equation}  
as demonstrated in \cite{Huang_2021}. Here, the matrix $A$ is defined as  
\begin{equation}
    A=(K+\lambda\mathds{1})^{-1}K(K+\lambda\mathds{1})^{-1},
\end{equation}  
where $K$ represents the kernel matrix and $\lambda$ is the regularization parameter. In Figure \ref{Fig:gen_bound}, we plot the first term of $\epsilon_{\mathrm{gen}}$, referred to as the generalization error bound, as a function of the hyperparameter $\lambda$ for $n=8$ qubits. When $\lambda=0$, the training error is zero, which corresponds to the maximum generalization error.  

As shown in Figure \ref{Fig:gen_bound}, employing the neural approach results in a lower generalization error bound compared to a problem-agnostic embedding with randomly chosen parameters. Furthermore, we observe that PQKs exhibit a lower generalization error bound overall. The difference between random and neural approaches is more pronounced in PQKs, whereas in EQKs, the advantage of the neural approach over the random one is less significant. These results represent the mean of five independent experiments conducted on the Fashion MNIST dataset with $M=50$ training samples.

\section{Conclusions}
Building meaningful quantum machine learning models requires designing problem-inspired quantum embeddings. Neural quantum kernels enable the encoding of problem-specific information into quantum kernel methods by training a neural network, offering a more efficient alternative to previous approaches that required constructing the kernel matrix at every training step. Through our proposed method for scaling QNNs, we demonstrate how they can effectively generate powerful neural quantum kernels capable of competing with well-optimized classical machine learning algorithms. Specifically, we introduce neural EQKs with two distinct constructions and a structured approach for designing neural PQKs. Our results highlight the advantages of neural quantum kernels over problem-agnostic embeddings in terms of performance, trainability, and generalization error. While neural EQKs achieve higher accuracy, neural PQKs exhibit greater robustness in noisy scenarios, as well as improved trainability and generalization. These findings pave the way for developing scalable and resilient quantum kernel models, with potential extensions to quantum-inspired methods.

\begin{acknowledgements}
The authors would like to thank Maria Schuld for her valuable comments which led to a more solid presentation of the results and Sofiene Yerbi and Adrián Pérez-Salinas for their worthy insights during QTML 2023. The 
authors acknowledge financial support from OpenSuperQ+100 (Grant No. 101113946) 
of the EU Flagship on Quantum Technologies, as well as from the EU FET-Open project 
EPIQUS (Grant No. 899368), also from Project Grant No. PID2021-125823NA-I00 595 and 
Spanish Ramón y Cajal Grant No. RYC-2020-030503-I funded by MCIN/AEI/10.13039/501100011033 and by “ERDF A way of making Europe” and “ERDF 
Invest in your Future,” this project has also received support from the Spanish Ministry for Digital Transformation and of Civil Service of the Spanish Government through the QUANTUM ENIA project call - Quantum Spain, EU through the Recovery, Transformation and Resilience Plan – NextGenerationEU within the framework of the Digital Spain 2026., and by the EU through the Recovery, Transformation and Resilience Plan – NextGenerationEU within the framework of the Digital Spain 2026 Agenda, we acknowledge funding from Basque Government through Grant No. IT1470-22 and the IKUR 
Strategy under the collaboration agreement between Ikerbasque Foundation and BCAM 
on behalf of the Department of Education of the Basque Government. PR and YB acknowledge financial support from the CDTI within the Misiones 2021 program and the Ministry of Science and Innovation under the Recovery, Transformation and Resilience Plan—Next Generation EU under the project “CUCO: Quantum Computing and its Application to Strategic Industries”. Y.B. acknowledges ayudas Ramon y Cajal (RYC2023-042699-I).

\end{acknowledgements}

\begin{appendix}
\begin{figure*}[t!]
\centering
\includegraphics[width=\textwidth]{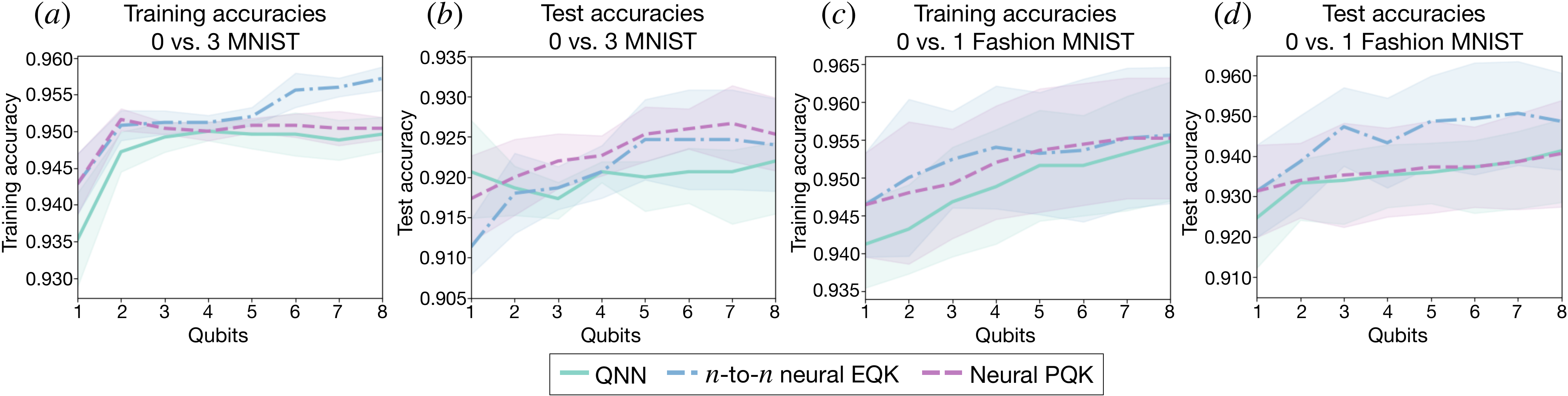}
\caption{Mean training ($(a)$ and $(c)$) and test accuracies ($(b)$ and $(d)$) using the proposed QNN construction and neural quantum kernels. We present experiments distinguishing between the digits 0 and 3 from the MNIST dataset ($(a)$ and $(b)$) and between t-shirt/top ($\hat{y}=0$) and trousers ($\hat{y}=1$) from the Fashion MNIST dataset ($(c)$ and $(d)$). The results are averaged over 5 runs, using 500 training points and 300 test points in each run.}
\label{Fig:extra}
\end{figure*}

\section{Training EQKs}
One of the main drawbacks of quantum kernel methods is determining the best kernel which depends on the specific problem. In cases where no prior knowledge about the input data is available, approaches have been developed to construct parametrized embedding quantum kernels that are trained for a specific task. These kernel training methods are based on multiple kernel learning \cite{vedaie2020quantum, ghukasyan2023quantumclassical} and kernel target alignment \cite{Hubregtsen_2022}.

\subsection{Multiple kernel learning}
Multiple kernel learning methods involve the creation of a combined kernel
\begin{equation}
k_{\boldsymbol{\varphi},\boldsymbol{\omega}}(\boldsymbol{x}_i,\boldsymbol{x}_j)=f_{\boldsymbol{\omega}}\left(\{k_{\boldsymbol{\varphi}}^{(r)}(\boldsymbol{x}_i,\boldsymbol{x}_j)\}_{r=1}^R\right),
\end{equation}
where $k_{\boldsymbol{\varphi}}^{(r)}$ represents a set of parametrized embedding quantum kernels, and $f_{\boldsymbol{\omega}}$ is a combination function. Two common choices for this function result in either a linear kernel combination
\begin{equation}
k_{\boldsymbol{\varphi},\boldsymbol{\omega}}(\boldsymbol{x}_i,\boldsymbol{x}_j)=\sum_{r=1}^R\omega_r k_{\boldsymbol{\varphi}}^{(r)}(\boldsymbol{x}_i,\boldsymbol{x}_j),
\end{equation}
or a multiplicative kernel combination
\begin{equation}
k_{\boldsymbol{\varphi}}(\boldsymbol{x}_i,\boldsymbol{x}_j)=\prod_{r=1}^R k_{\boldsymbol{\varphi}}^{(r)}(\boldsymbol{x}_i,\boldsymbol{x}_j).
\end{equation}
When it comes to choosing model parameters, in Ref.~\cite{vedaie2020quantum} they use an empirical risk function to minimize, while in Ref.~\cite{ghukasyan2023quantumclassical} the authors opt for a convex minimization problem. However, in both constructions a complete kernel matrix is computed at each optimization step.

\subsection{Kernel target alignment}
Training kernels using kernel target alignment is based on the similarity measure between two kernel matrices, denoted as $K_A$ and $K_B$. This similarity measure, known as kernel alignment \cite{holmes2006innovations}, is defined as
\begin{equation}
KA(K_A,K_B) = \frac{\mathrm{tr}(K_A K_B)}{\sqrt{\mathrm{tr}(K_A^2) \cdot \mathrm{tr}(K_B^2)}}.
\end{equation}
and corresponds to the cosine of the angle between kernel matrices if we see them as vectors in the space of matrices with the Hilbert-Schmidt inner product. The use of kernel alignment to train kernels is by defining the ideal kernel matrix $K^*$ whose entries are $k^*_{ij}=k^*(\boldsymbol{x}_i,\boldsymbol{x}_j)=y_i y_j$. Given a parametrized quantum embedding kernel $k_{\boldsymbol{\theta}}(\boldsymbol{x}_i,\boldsymbol{x}_j)=|\langle \phi_{\boldsymbol{\theta}}(\boldsymbol{x}_i)|\phi_{\boldsymbol{\theta}}(\boldsymbol{x}_j)\rangle|^2$ whose kernel matrix is named as $K_{\boldsymbol{\theta}}$, the kernel-target alignment is defined as the kernel alignment between $K$ and the ideal kernel, i.e.
\begin{equation}
    TA(K_{\boldsymbol{\theta}})=KA(K_{\boldsymbol{\theta}},K^*)=\frac{\sum_{ij}y_i y_j\; k_{\boldsymbol{\theta}}(\boldsymbol{x}_i,\boldsymbol{x}_j)}{M\sqrt{\sum_{ij}k_{\boldsymbol{\theta}}(\boldsymbol{x}_i,\boldsymbol{x}_j)^2}},
\end{equation}
where we used that $y_i^2=1$ independently from the label and $M$ is the number of training points. Thus, authors in Ref.~\cite{Hubregtsen_2022} use this quantity as cost function which is minimized using gradient descent over the $\boldsymbol{\theta}$ parameters. This strategy is closely related to the approach outlined in Ref.~\cite{lloyd2020quantum}, where the authors explore the construction of quantum feature maps with the aim of maximizing the separation between different classes in Hilbert space. However, in the case of the kernel target alignment strategy, similar to multiple kernel learning, the construction of the kernel matrix (or at least a subset of it) at each training step is required.

\section{Data re-uploading}\label{reuploading}
To construct a data re-uploading QNN for binary classification, each label is associated with a unique quantum state, aiming to maximize the separation in the Bloch sphere. For the single qubit quantum classifier, the labels +1 and -1 are represented by the computational basis states $|0\rangle$ and $|1\rangle$, respectively. The objective is to appropriately tune the parameters $\{\boldsymbol{\theta}_l\}_{l=1}^L$ that define the state
\begin{equation}
|\phi_{\boldsymbol{\theta}}(\boldsymbol{x}_i)\rangle=U(\boldsymbol{\theta}_L )\;U(\boldsymbol{x}) \dots U(\boldsymbol{\theta}_1)\;U(\boldsymbol{x})|0\rangle,
\end{equation}
to rotate the data points of the same class close to their corresponding label state. To achieve this, we use the fidelity cost function
\begin{equation}\label{cost_function}
    f_\mathrm{cost}=\frac{1}{M}\sum_{i=1}^M \Bigl(1-|\langle \phi^i_l|\phi_{\boldsymbol{\theta}}(\boldsymbol{x}_i)\rangle|^2\Bigr),
\end{equation}
where $|\phi_l^i\rangle$ represents the correct label state for the data point $\boldsymbol{x}_i$. This optimization process occurs on a classical processor. Once the model is trained, the quantum circuit is applied to a test data point $\boldsymbol{x}_t$, and the probability of obtaining one of the label states is measured. If this probability surpasses a certain threshold (set as 1/2 in this case), the data point is classified into the corresponding label state class. Formally, the decision rule can be expressed as
\begin{equation}\label{main_decisionrule}
    \hat{y}[\boldsymbol{x}_t] = \begin{cases}
+1 & \text{if } |\langle 0|\phi(\boldsymbol{x}_t)\rangle|^2 \geq 1/2, \\
-1 & \text{if } |\langle 0|\phi(\boldsymbol{x}_t)\rangle|^2 < 1/2.
\end{cases}
\end{equation}
When consider a multi-qubit architecture, the idea is the same but we need to properly choose the corresponding label states. In this work, for a $n$-qubit QNN we consider as label states the ones defined by the projectors $|0\rangle\langle0|\otimes \mathds{1}^{(n-1)}$ and $|1\rangle\langle1|\otimes \mathds{1}^{(n-1)}$, which corresponds to a local measurement in the first qubit.

\begin{figure*}[t!]
\centering
\includegraphics[width=\textwidth]{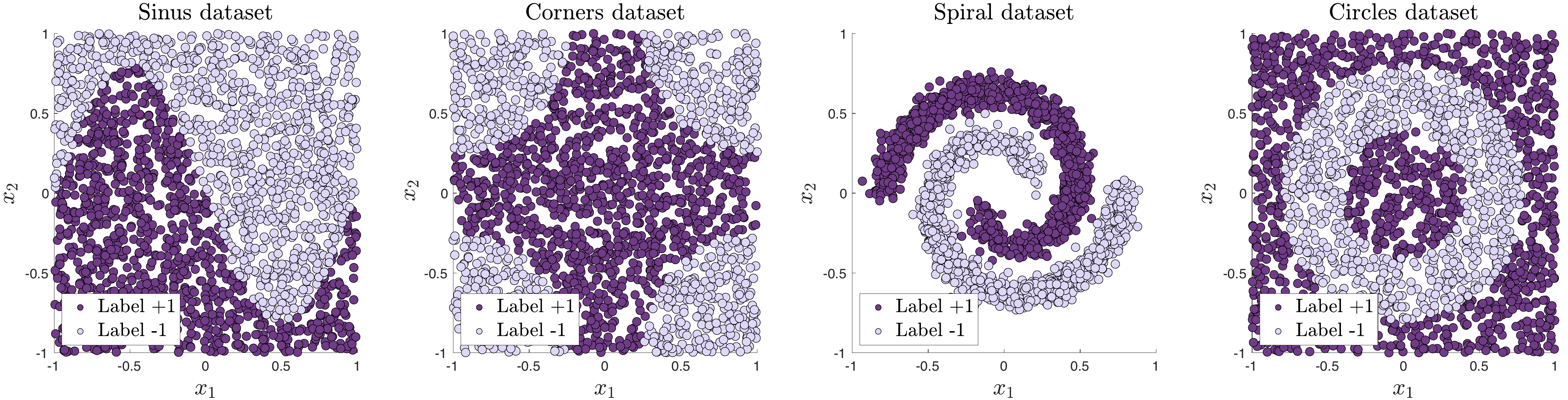}
\caption{ From left to right: sinus, corners, spiral and circles datasets generated and used in this work. These figures are generated with 2000 data points.}
\label{Fig:datasets}
\end{figure*}

\section{Hyperplane defined in the Bloch sphere}\label{hyperplane}
Here we demonstrate that when the QNN training is perfect, constructing the kernel becomes redundant since the decision measurement remains unchanged. 

The objective of the data re-uploading QNN is to map all points with label +1 to the $|0\rangle$ state on the Bloch sphere, while mapping points with label -1 to the $|1\rangle$ state. In an ideal scenario, all +1 points would be rotated to the $|0\rangle$ state and all -1 points to the $|1\rangle$ state. Using this feature map to construct the EQK, all points would be support vectors associated to the same Lagrange multiplier $\alpha$. The SVM generates a separating hyperplane defined by the equation
\begin{equation}\label{hyper_svm_appendix}
    \sum_{i\in\mathrm{SV}} \alpha_i\;y_i\; k(\boldsymbol{x}_i,\boldsymbol{x})+b=0.
\end{equation}
In this scenario, assuming a balanced dataset translates to an equal number of support vectors, denoted as $N_{\mathrm{SV}}$, for each class, and results in $b=0$ due to symmetry. Working with this equation by considering the sum of support vectors separately for the +1 class ($\mathrm{SV}_{+1}$) and -1 class ($\mathrm{SV}_{-1}$), we obtain
\begin{equation}
    \begin{split}
        \sum_{i\in \mathrm{SV}_{+1}}\alpha_i\;y_i\;|\langle &\phi(\boldsymbol{x}_i)|\phi(\boldsymbol{x})\rangle|^2+\sum_{i\in \mathrm{SV}_{+1}}\alpha_i\;y_i\;|\langle \phi(\boldsymbol{x}_i)|\phi(\boldsymbol{x})\rangle|^2\\
        &=N_\mathrm{SV}\alpha\; |\langle 0|\phi(\boldsymbol{x}\rangle |^2-N_\mathrm{SV}\alpha\; |\langle 1|\phi(\boldsymbol{x}\rangle |^2=0,
    \end{split}
\end{equation}
which is equivalent to the hyperplane defined by
\begin{equation}\label{ap:plane}
    |\langle 0|\phi(\boldsymbol{x}\rangle |^2=|\langle 1|\phi(\boldsymbol{x}\rangle |^2,
\end{equation}
mirroring the decision boundary of the QNN part. Therefore, in the case of a perfect training, the SVM becomes redundant. Even if the points are not perfectly mapped to their label states and the separating hyperplane of the SVM is not exactly the one from Eq.~\ref{ap:plane}, as long as all points are correctly classified by the QNN, the SVM will yield the same results. Thus, the SVM part is only meaningful for the single-qubit QNN to construct a single-qubit EQK case when the QNN training is suboptimal and requires to adjust the measurement of the decision boundary.

\vspace{1cm}

\section{Additional numerical simulations on real datasets}
We present additional numerical experiments on two different datasets, corresponding to ideal simulations using our proposed QNN, and two types of neural quantum kernels: the n-to-n neural EQK and the neural PQK, as discussed in the main text. The results are shown in Figure \ref{Fig:extra}, displaying both the mean and standard deviation of training and test accuracies over five different runs, using 500 training samples and 300 test samples. The two classification tasks considered are distinguishing between the digits 0 and 3 from the MNIST dataset (shown in Figures \ref{Fig:extra} $(a)$ and $(b)$), and distinguishing between t-shirt/top and trousers from the Fashion MNIST dataset (shown in Figures \ref{Fig:extra} $(c)$ and $(d)$). The conclusions drawn from these results align with those in the main text: adding qubits improves the performance, and kernel methods constructed from the QNN consistently outperform other approaches. Which of the two proposed neural quantum kernels performs better depends on the classification task at hand. While the neural PQK achieves better test accuracy for the MNIST dataset, the neural EQK performs better for the Fashion MNIST dataset.

\section{Additional numerical simulations on toy datasets}
\subsection{Datasets}
For additional numerical simulations we utilized four artificial datasets depicted in Figure \ref{Fig:datasets} for evaluating the performance of the proposed neural EQKs. Here we explain how they were created:
\begin{itemize}
    \item \textbf{Sinus Dataset}: The sinus dataset is defined by a sinusoidal function, specifically $f(x_1) = -0.8\;\sin(\pi x_1)$. Points located above this sinusoidal curve are categorized as class -1, while points below it are assigned to class +1.
    \item \textbf{Corners Dataset}: The corners dataset comprises four quarters of a circumference with a radius of 0.75, positioned at the corners of a square. Points located inside these circular regions are labeled as class -1, while points outside are classified as class +1.
    \item \textbf{Spiral Dataset}: The spiral dataset features two spirals formed by points arranged along a trajectory defined by polar coordinates. The first spiral, denoted as class +1, originates at the origin (0, 0) and spirals outward in a counter-clockwise direction, forming a curve. The second spiral, labeled class -1, mirrors the first spiral but spirals inward in a clockwise direction. These spirals are generated by varying the polar angle, selected randomly to create the data points. The radial distance from the origin for each point depends on the angle, creating the characteristic spiral shape. Noise is added to the data points by introducing random perturbations to ensure they do not align perfectly along the spirals.
    \item \textbf{Circles Dataset}: The circles dataset is created using two concentric circles that define an annular region. The inner circle has a radius of $\sqrt{2/\pi}$, while the outer circle has a radius of $0.5\sqrt{2/\pi}$. Data points located within the annular region are labeled as -1, while those outside the region are labeled as +1.
\end{itemize}

\begin{figure*}[t!]
\centering
\includegraphics[width=\textwidth]{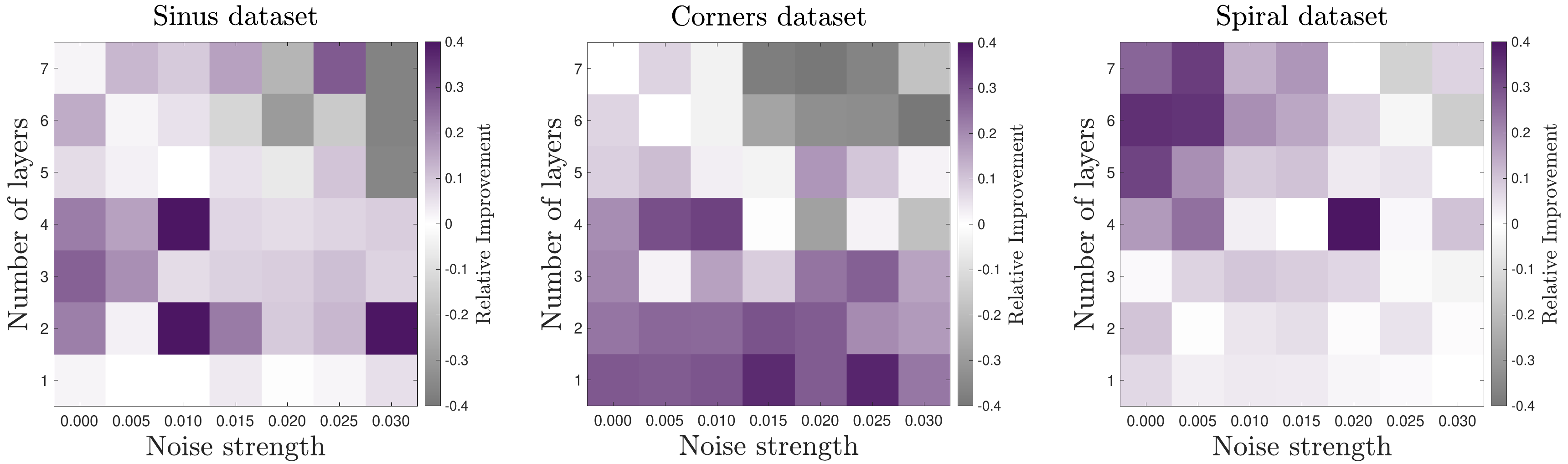}
\caption{Relative improvement, as a function of the number of layers and noise strength, comparing the single qubit QNN accuracy performance with the combined protocol utilizing a two-qubit embedding kernel. Results are shown for the sinus, corners, and spiral datasets.}
\label{Fig:noisy}
\end{figure*}
\subsection{Additional noisy simulations}\label{noisy_sim}
We have introduced a combined protocol for binary classification. It is worth noting that the kernel estimation part involves the utilization of a quantum circuit with twice the depth of the QNN part. In practical implementations on current noisy intermediate-scale quantum devices, the consideration of a larger circuit warrants careful attention. This is due to the susceptibility of larger circuits to elevated noise levels, potentially affecting the overall performance.

As previously mentioned, increasing the number of layers in the model enhances its expressivity. However, when constructing the kernel using a high number of layers, noise can have a detrimental impact on the performance, resulting in poorer results for the combined protocol compared to using only the QNN. Therefore, in this section, our objective is to perform simulations to visualize the trade-off between the number of layers in the QNN and using the combined protocol.

To characterize the noise incorporated into these simulations, we employ the operator sum representation of the noise channel $\varepsilon$ acting on a quantum state $\rho$,
\begin{equation}
    \varepsilon(\rho) = \sum_i K_i \rho K_i^\dagger,
\end{equation}
where $K_i$ represents the respective Kraus operators. In our simulations, we account for two single-qubit noise channels that are applied after each quantum gate. Firstly, we consider amplitude damping error which is described by the Kraus operators
\begin{align}
    K_0&=
    \begin{pmatrix}
        1 & 0\\
        0 & \sqrt{1-\gamma}
    \end{pmatrix},\\
    K_1&=
    \begin{pmatrix}
        0 & \sqrt{\gamma} \\
        0 & 0
    \end{pmatrix},
\end{align}
Here, $\gamma\in [0,1]$ is the amplitude damping probability, which can be defined as $\gamma = 1 - e^{-\Delta t / T_1}$, with $T_1$ representing the thermal relaxation time and $\Delta t$ denoting the duration of application of a quantum gate. The second noise source under consideration is the phase flip error, which can be described by the following Kraus operators
\begin{align}
    K_0 &= \sqrt{1-\alpha}
    \begin{pmatrix}
        1 & 0 \\
        0 & 1
    \end{pmatrix}, \\
    K_1 &= \sqrt{\alpha}
    \begin{pmatrix}
        1 & 0 \\
        0 & -1
    \end{pmatrix}.
\end{align}In this case, $\alpha\in [0,1]$ represents the probability of a phase flip error, which can be defined as $2\alpha = 1 - e^{-\Delta t / (2T_2)}$, with $T_2$ denoting the spin-spin relaxation time or dephasing time. The value of $\alpha$ also falls within the interval $[0, 1]$.

The current state-of-the-art experimental values for the noise parameters of a superconducting quantum processor are as follows: $T_1$ falls within the range of $50-150 \; \mu s$, $T_2$ is in the range of $25-75 \; \mu s$, and $\Delta t$ varies from $10-50 \; ns$. It is important to note that noise becomes more pronounced when $T_1$ and $T_2$ decrease and/or when $\Delta t$ increases. To explore the scenario with the worst noise, we consider the extreme values within these ranges, leading to $\gamma^* = 0.001$ and $\alpha^* = 0.0005$.

In our simulations, we simplify the noise characterization by defining a noise strength parameter $\tau\equiv\alpha = \gamma$ for the sake of simplicity. We choose the single-qubit data re-uploading quantum neural network (QNN) as the kernel selection part. To observe a transition where it is no longer advantageous to include the support vector machine (SVM) part, we explore values of $\tau$ ranging from $0.005$ to $0.030$ in steps of $0.005$, while also including the noiseless case where $\tau = 0$. These values correspond to examining the range $5\gamma^* \leq \gamma \leq 30\gamma^*$ and $10\alpha^* \leq \alpha \leq 60\alpha^*$. Notably, even in these highly adverse conditions, which are significantly worse than those of current real hardware quantum devices, we find that the combined protocol remains suitable. This conclusion aligns with expectations since the protocol utilizes only a small number of qubits and quantum gates.
\begin{figure*}[t!]
\centering
\includegraphics[width=\textwidth]{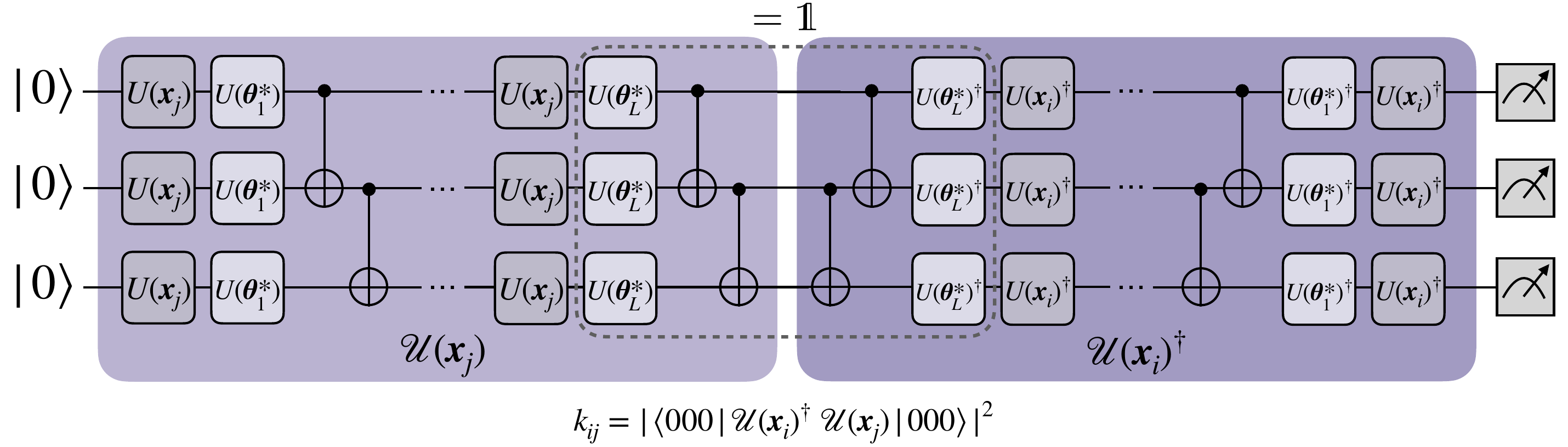}
\caption{Explicit quantum circuit for constructing the $1$-to-$3$ EQK with cascade of $\mathrm{CNOT}$ gates as source of entanglement. The parameters $\boldsymbol{\theta}^*$ are the ones obtained during the training of the single-qubit QNN. Note that the parameters of the last layer are irrelevant.}
\label{Fig:multi_QNN}
\end{figure*}
In Figure~\ref{Fig:noisy}, we present the relative improvement, denoted as
\begin{equation}
\text{Relative improvement}=\frac{\mathrm{Acc}_{\mathrm{QNN+SVM}}-\mathrm{Acc}_{\mathrm{QNN}}}{\mathrm{Acc}_{\mathrm{QNN}}},
\end{equation}
plotted as a function of the noise strength parameter $\tau$ and the number of layers $L$. The architecture employed corresponds to the $1$-to-$2$ case, utilizing $\mathrm{CNOT}$ gates for entanglement between layers. For QNN training in this instance, we limit it to two epochs with a learning rate of 0.05. The objective is to demonstrate that the resulting EQK is not highly dependent on the training specifics of the corresponding QNN and even a sub-optimal QNN training can lead to a powerful EQK for the specific task.

The relative improvements vary depending on the dataset under consideration and range from $-40\%$ to $40\%$. As expected, the worst performance of the combined protocol is observed at higher noise levels and for a greater number of layers, corresponding to a longer-depth quantum circuit. It is crucial to emphasize once more that these high noise strength values significantly surpass the noise parameters of current quantum devices. Therefore, even though the combined protocol necessitates running a double-depth circuit with more qubits, considering the noise model described earlier, it remains suitable for actual quantum devices.

\subsection{Additional ideal simulations}\label{numerical_experiments}
In the main text we provide numerical results for the Fashion MNIST dataset. Here we provide the tables of more numerical experiments for the neural EQKs construction using the same hyperparameters: learning rate of 0.05 and 30 epochs for the $n=1$ QNN and learning rate of 0.005 and 10 epochs for $n>1$. Here we provide more results for different toy datasets and considering also CZ as source of entanglement. All accuracies refer to test accuracies.

\begin{table}[h!]
    \centering
    \begin{tabular}{c||c||c||c}
    \hline
    \hline
    \textbf{Dataset} & \textbf{QNN accuracy} & \textbf{EQK type} & \textbf{EQK accuracy} \\
    \hline
      \hline
    Sinus & 0.890 & $3q$ $\mathrm{CNOT}$ & 0.960 \\
    Sinus &  & $3q$ $\mathrm{CZ}$ & 0.948 \\
    \hline
    Corners & 0.886 & $3q$ $\mathrm{CNOT}$ &  0.954 \\
    Corners &  & $3q$ $\mathrm{CZ}$ & 0.940 \\
    \hline
    Spiral & 0.800 & $3q$ $\mathrm{CNOT}$ & 0.994 \\
    Spiral &  & $3q$ $\mathrm{CZ}$ & 0.866 \\
    \hline
    Circles & 0.698 & $3q$ $\mathrm{CNOT}$ & 0.866 \\
    Circles &  & $3q$ $\mathrm{CZ}$ & 0.864 \\
    \hline
  \end{tabular}
    \caption{Numerical results for the $1$-to-$3$ architecture, considering two types of entangling gates ($\mathrm{CNOT}$ and $\mathrm{CZ}$), on four distinct datasets.}
    \label{tab:1_to_2,3}
\end{table}

In Table \ref{tab:1_to_2,3}, the outcomes demonstrate striking similarity when introducing entanglement through $\mathrm{CZ}$ gates or $\mathrm{CNOT}$ gates, with the exception of the spiral dataset. Given the ambiguity surrounding the method of entanglement introduction, one might contemplate utilizing controlled rotations with random parameters as a potential source of entanglement. Notably, it is observed that even starting from a single-qubit QNN achieving accuracies of less than $90\%$, the construction of EQKs yields accuracies surpassing $95\%$.

\begin{table}[h!]
    \centering
    \begin{tabular}{c||c|c|c|c|c|c|c|c|c}
    \hline
    \hline
    \textbf{Dataset} & \textbf{n=2} & \textbf{n=3} & \textbf{n=4} & \textbf{n=5} & \textbf{n=6} & \textbf{n=7} & \textbf{n=8} & \textbf{n=9} & \textbf{n=10} \\
    \hline
      \hline
    Corners & 0.890 & 0.954 & 0.974 & 0.978 & 0.982 & 0.980 & 0.978 & 0.978 & 0.978 \\
    \hline
    Circles & 0.832 & 0.866 & 0.950 & 0.970 & 0.974 & 0.970 & 0.978 & 0.966 & 0.962 \\
    \hline
  \end{tabular}
    \caption{Numerical results for the $1$-to-$n$ architecture for up to $n=10$ qubits.}
    \label{tab:1_to_n}
\end{table}

In the $1$-to-$n$ construction presented in Table \ref{tab:1_to_n} for the corners and circles datasets, we observe that the accuracy rises rapidly as qubits are added, reaching a peak at $n=4$. Subsequently, the accuracy plateaus, attaining a maximum at $n=6$ for the corners dataset and $n=8$ for the circles dataset. After reaching these points, the accuracy gradually decreases with additional qubits.

\begin{table}[h!]
    \centering
    \begin{tabular}{c||c|c|c|c|c|c|c|c}
    \hline
    \hline
    \textbf{Dataset} & \textbf{n=1} & \textbf{n=2} & \textbf{n=3} & \textbf{n=4} & \textbf{n=5} & \textbf{n=6} & \textbf{n=7} & \textbf{n=8}  \\
    \hline
      \hline
    Corners (QNN) & 0.886 & 0.934 & 0.948 & 0.952 & 0.954 & 0.956 & 0.960 & 0.962  \\
    \hline
    Corners (EQK) & 0.898 & 0.948 & 0.960 & 0.966 & 0.970 & 0.972 & 0.974 & 0.974 \\
    \hline
    \hline
    Circles (QNN) & 0.698 & 0.792 & 0.820 & 0.858 & 0.934 & 0.944 & 0.944 & 0.946 \\
    \hline
    Circles (EQK) & 0.796 & 0.820 & 0.906 & 0.968 & 0.974 & 0.974 & 0.976 & 0.980 \\
    \hline
  \end{tabular}
    \caption{Numerical results for the $n$-to-$n$ architecture for up to $n=8$ qubits. These results are depicted in the main text.}
    \label{tab:n_to_n}
\end{table}

In the $n$-to-$n$ approach, as presented in Table \ref{tab:n_to_n}, the accuracy consistently increases with the addition of qubits for both the QNN and the EQK. Additionally, as discussed in the main text, the EQK consistently outperforms the corresponding QNN architecture for the same value of $n$.

\begin{table}[h!]
    \centering
    \begin{tabular}{c||c||c||c}
    \hline
    \hline
    \textbf{Dataset} & \textbf{Layers} & \textbf{QNN accuracy} & \textbf{EQK accuracy} \\
    \hline
      \hline
    Sinus & $L=5$ & 0.948 & 0.970 \\
    Sinus & $L=6$ & 0.956 & 0.964 \\
    Sinus & $L=7$ & 0.966 & 0.972 \\
    Sinus & $L=8$ & 0.958 & 0.964 \\
    \hline
    Corners & $L=5$ & 0.948 & 0.970 \\
    Corners & $L=6$ & 0.936 & 0.950 \\
    Corners & $L=7$ & 0.934 & 0.948 \\
    Corners & $L=8$ & 0.916 & 0.920 \\
    \hline
    Spiral & $L=5$ & 0.952 & 0.996 \\
    Spiral & $L=6$ & 0.974 & 0.998 \\
    Spiral & $L=7$ & 0.978 & 1.000 \\
    Spiral & $L=8$ & 0.980 & 0.998 \\
    \hline
    Circles & $L=5$ & 0.786 & 0.812 \\
    Circles & $L=6$ & 0.808 & 0.814 \\
    Circles & $L=7$ & 0.792 & 0.820 \\
    Circles & $L=8$ & 0.844 & 0.902 \\
    \hline
  \end{tabular}
    \caption{Numerical results for the $2$-to-$2$ architecture for different number of layers. For the experiments in the main text we choose $L=7$.}
    \label{tab:layers}
\end{table}

Finally, in Table \ref{tab:layers}, we present results for the four datasets considering different numbers of layers for the $n$-to-$n$ architecture with $n=2$. Adding layers increases the expressivity of the QNN but does not guarantee better accuracies, as we can observe. Again, we see that the EQK consistently outperforms the QNN.

\end{appendix}

\clearpage
\bibliography{main}
\end{document}